\documentclass[12pt,a4paper,final]{iopart}

\usepackage{iopams}  
\usepackage{graphicx}
\usepackage[breaklinks=true,colorlinks=true,linkcolor=blue,urlcolor=blue,citecolor=blue]{hyperref}
\usepackage{amsfonts}

\expandafter\let\csname equation*\endcsname\relax
\expandafter\let\csname endequation*\endcsname\relax
\usepackage{amsmath}

\usepackage{amssymb}
\usepackage{bm}
\usepackage{dcolumn}
\usepackage{graphicx}
\usepackage{graphics}
\usepackage[latin1]{inputenc}
\usepackage{latexsym}
\usepackage{rotating}
\usepackage{hyperref}
\usepackage[all]{hypcap} 
\usepackage{xspace} 
\usepackage[usenames,dvipsnames]{color}
\usepackage{mathrsfs}
\usepackage{subfigure}

\usepackage{ulem}
\usepackage{outlines}
\usepackage{enumitem}
\setenumerate[1]{label=\Roman*.}
\setenumerate[2]{label=\Alph*.}
\setenumerate[3]{label=\roman*.}
\setenumerate[4]{label=\alph*.}

\definecolor {darkgreen}{rgb}{0.2,0.7,0.2}

\newcommand\be{\begin{equation}}
\newcommand\ba{\begin{eqnarray}}
\newcommand\ee{\end{equation}}
\newcommand\ea{\end{eqnarray}}

\newcommand\bw{\begin{widetext}}
\newcommand\ew{\end{widetext}}

\newcommand{\nn}{\nonumber}

\newcommand{\GR}{{\mbox{\tiny GR}}}

\newcommand{\Pl}{{\mbox{\tiny Pl}}}

\begin{document}
\title{Eccentric Gravitational Wave Bursts in the Post-Newtonian Formalism}

\author{Nicholas Loutrel}
\address{eXtreme Gravity Institute, Department of Physics, Montana State University, Bozeman, MT 59717, USA.}
\ead{nicholas.loutrel@montana.edu}

\author{Nicol\'as Yunes}
\address{eXtreme Gravity Institute, Department of Physics, Montana State University, Bozeman, MT 59717, USA.}
\ead{nicolas.yunes@montana.edu}

\date{\today}

\begin{abstract} 

The detection of GW150914 by ground based gravitational wave observatories has brought about a new era in astrophysics. At optimal sensitivity, these observatories are expected to detect several events each year, with one or two of these occurring with non-negligible eccentricity. Such eccentric binaries will emit bursts of gravitational radiation during every pericenter passage, where orbital velocities can reach greater than ten percent the speed of light. As a result, such binaries may prove to be powerful probes of extreme gravitational physics and astrophysics. A promising method of achieving detection of such binaries is through power stacking, where the power in each burst is added up in time-frequency space. This detection strategy requires a theoretical prior of where the bursts will occur in time and frequency so that one knows where to search for successive bursts. We here present a generic post-Newtonian formalism for constructing such time-frequency model priors at generic post-Newtonian order. We apply our formalism to generate a burst model at third post-Newtonian order, making it potentially the most accurate, fully analytic model to date.

\end{abstract}

\pacs{04.30.-w,04.25.-g,04.25.Nx}


\maketitle


\section{Introduction}
\label{intro}

Over the past several years, studies have shown that exotic formation channels could lead to a population of highly eccentric compact binaries whose gravitational wave (GW) emission would be in the band of current ground-based detectors~\cite{2009MNRAS.395.2127O, Lee:2009ca,Wen:2002km,Kushnir:2013hpa,2013PhRvL.111f1106S,Antognini:2013lpa,Naoz:2012bx,Antonini:2013tea,Antonini:2015zsa, East:2012xq}. One such formation channel is dynamical captures in dense stellar environments, such as globular clusters and galactic nuclei~\cite{2009MNRAS.395.2127O, Lee:2009ca, East:2012xq}. In such dense environments, compact objects initially on hyperbolic trajectories can become bound after passing through closest approach due to GW emission or tidal interactions, with the subsequent bound binary having high eccentricity ($ e \sim 1$). On the other hand, the Kozai-Lidov mechanism~\cite{Wen:2002km,Naoz:2011mb,Kushnir:2013hpa,2013PhRvL.111f1106S,Antognini:2013lpa,Naoz:2012bx,Antonini:2013tea,Antonini:2015zsa, VanLandingham:2016ccd}, and other three-body interactions~\cite{Naoz:2012bx} in hierarchical triple systems, can induce resonances that drive the inner binary to the parabolic limit.

Although expected to be rare~\cite{Abadie:2010cf}, eccentric binaries could prove to be powerful probes of astrophysical dynamics. Event rates of eccentric inspirals due to dynamical captures have wide error bars, typically around two orders of magnitude~\cite{2009MNRAS.395.2127O}. Such large error is largely due to the unconstrained populations of compact objects within dense environments~\cite{Abadie:2010cf}. Likewise, the tightening of binaries due to three-body interactions in galactic nuclei are similarly uncertain~\cite{Miller:2008yw}. Hence, the detection of GWs from eccentric inspirals would provide information about the mass function of black holes (BHs) and neutron stars (NSs) in these environments, allowing us to probe astrophysics that has proven difficult to extract from electromagnetic observations~\cite{O'Leary:2007qa, Zevin:2017evb, Rodriguez:2016geq, Stevenson:2017dlk}.

Another promising area of interest for eccentric binaries is testing Einstein's theory of General Relativity (GR). For highly eccentric binaries, the distance of closest approach can be small relative to the semi-major axis of the orbit, leading to systems with pericenter velocities greater than $10\%$ the speed of light. At such high velocities, the GW luminosity \emph{in each burst} will typically be $\sim (10^{-4}-10^{-3}) L_{\Pl}$, where $L_{\Pl}$ is the Planck luminosity. For comparison, a BH-BH, quasi-circular binary emits radiation at $\lesssim10^{-5} L_{\Pl}$ during the early phase of inspiral, increasing rapidly close to merger and eventually reaching $\sim 10^{-2} L_{\Pl}$ only at merger. As such, the GWs from eccentric binaries could capture effects from the extreme gravity regime (i.e.~where gravity is dynamical, strong and non-linear) during many pericenter passages~\cite{Loutrel:2014vja}.

These GWs may be detected by current and upcoming ground-based GW detectors, and hence the study of GWs from eccentric binaries has never been more urgent. The advanced Laser Interferometer Gravitational Wave Observatory (aLIGO)~\cite{Harry:2010zz} has already achieved the first detection of GWs with the event GW150914~\cite{Abbott:2016blz}. The advanced VIRGO Interferometer (aVIRGO)~\cite{TheVirgo:2014hva} will be coming online in 2016-2017, with additional detections expected during this period. Plausible estimates of the event rates of the inspiral of compact object binaries predict that these detectors could see $\sim 5-10$ events per year in the very near future~\cite{Abadie:2010cf,Abbott:2016nhf,Abbott:2016drs}. Based on our current knowledge of the formation channels of eccentric compact binaries~\cite{2009MNRAS.395.2127O, Lee:2009ca,Wen:2002km,Kushnir:2013hpa,2013PhRvL.111f1106S,Antognini:2013lpa,Naoz:2012bx,Antonini:2013tea,Naoz:2012bx,Antonini:2015zsa}, we might expect that one or two of these events will enter the LIGO band with non-negligible orbital eccentricity. Looking toward the future, KAGRA~\cite{Uchiyama:2004vr,Somiya:2011np} in Japan is expected to come online by the end of this decade and LIGO India~\cite{Unnikrishnan:2013qwa} in the beginning of the next decade. Once these detectors are operational, the number of detected events per year will necessarily increase, thus increasing the probability of detecting eccentric inspirals.

The typical strategy for detecting well-understood GWs with ground-based interferometers is to use matched filtering~\cite{Brown:2012nn, Farr:2009pg, VanDenBroeck:2009gd, Ajith:2012az, Aasi:2012rja, Abadie:2011kd, Colaboration:2011np}. Effectively, a set of templates that best describe the signal buried in the data are used to extract the latter and estimate its physical parameters. This detection strategy hinges on having very accurate templates, as a small dephasing between the signal and a template can result in complete loss of detection~\cite{Sampson:2013jpa,Yunes:2009ke,Yagi:2013baa,Favata:2013rwa}. Hence, we must have a prior model of what to search for in the detector data output. For highly eccentric binaries, the GW emission resembles a set of discrete bursts somewhat localized in time and frequency. These bursts are centered around pericenter passage, where the orbital velocity is highest and where the binary spends the least amount of time, and thus very little GW power is contained within each individual burst. This issue alone makes matched filtering a rather impractical search strategy for such GWs, but it is further compounded by the fact that there are few, fully-analytic and accurate templates for highly eccentric binaries with which to perform matched filtering computationally efficiently~\cite{Hinder:2008kv, Huerta:2014eca}.

An alternative search strategy was presented in~\cite{Tai:2014bfa} using a power stacking method. Ideally, if one could register a set of bursts in a time-frequency decomposition of the data stream, one could then stack the power within each burst, thus creating an enhanced data product. For $N$ bursts with the same signal-to-noise ratio (SNR), the amplification in the SNR relative to the SNR in a single burst would scale as $N^{1/4}$. Although sub-optimal compared to matched filtering, the power stacking method is more robust to modeling errors and more efficient in detection of eccentric signals than current un-modeled burst methods~\cite{Tai:2014bfa}. Power stacking, however, still requires a model of where the bursts will occur in time-frequency space given some initial starting point, with which to construct a prior to search for successive bursts.

Such a burst model was developed in~\cite{Loutrel:2014vja} for tracking the bursts in time-frequency space. In general, a burst model is one that treats the bursts as N-dimensional objects in the detector's data stream and tracks the geometric centroid and volume of the bursts from one to the next as the binary inspirals. To do this,~\cite{Loutrel:2014vja} considered Keplerian orbits perturbed by quadrupolar gravitational radiation. We will refer to this model as \emph{Newtonian} in the sense that it is obtained from a fully relativistic model expanded to lowest, non-vanishing order about small orbital velocities and weak gravitational fields. The benefit of working in such a simplified scenario was that it allowed for the exploration of whether such burst signals could be used to test well-motivated deviations from GR with eccentric signals as a proof-of-concept.

The ultimate goal, of course, is to create a model that is as accurate as possible relative to the signals present in nature. The modeling of the coalescence of compact objects in full GR is an exceedingly difficult problem, which has only been solved numerically (predominantly for quasicircular binaries) in the passed several years. For eccentric binaries, numerical simulations in full GR are more computationally expensive, and thus, to obtain only a few orbits at the desired numerical accuracy requires much more computational time than that needed in the evolution of quasicircular binaries. At least for the moment, the pure numerical modeling of eccentric binaries over the last thousand orbits in full GR is currently an intractable problem. On the other hand, we could work to extend the Newtonian burst model by considering relativistic corrections to Newtonian dynamics.

The post-Newtonian (PN) formalism~\cite{Lorentz1937, Chandrasekhar:1965, Chandrasekhar:1969, Chandrasekhar:1970, Blanchet:2013haa,PW} allows for a systematic treatment of $v/c$ corrections to Newtonian dynamics, where $v$ is the orbital velocity and $c$ is the speed of light. For bound binaries, the orbital velocity is connected to the gravitational field strength through a Virial relation: a term of ${\cal{O}}(v^{2}/c^{2})$ is comparable to a term of ${\cal{O}}[GM/(R c^{2})]$, where $M$ and $R$ are the characteristic mass and orbital separation of the system, and $G$ is Newton's gravitational constant. Hence, the PN formalism for binaries is simultaneously a post-Minkowskian expansion, i.e.~it is both an expansion in $v/c \ll 1$ and an expansion in $GM/(Rc^2) \ll 1$. 

The PN formalism has had a wide range of success in the modeling of binaries and their GW emission. At present, the GW emission from quasicircular binaries has been calculated to 3.5 PN order\footnote{A term proportional to $(v/c)^{2N}$ relative to its controlling factor will be said to be of N PN order.}~\cite{Blanchet:2013haa} and to 4PN order in the effective one-body Hamiltonian~\cite{Damour:2014jta}. To leading order in the mass ratio, the radiation fluxes at spatial infinity are currently known to 22 PN order~\cite{Fujita:2012cm}, due to the formulaic nature of the calculation. The PN corrections to Newtonian dynamics for eccentric binaries have proven more difficult to calculate. Damour and Deruelle~\cite{zbMATH03938612,zbMATH04001537} found a Keplerian parameterization of the solution to the 1PN order equations of motion in terms of the eccentric anomaly $u$. This quasi-Keplerian (QK) representation was extended to 2PN order in~\cite{0264-9381-12-4-009} and to 3PN order in~\cite{Memmesheimer:2004cv}.

The QK parametrization must be enhanced to include dissipation if one wishes to obtain accurate waveform models. This parameterization is a purely conservative representation of the orbital motion of eccentric binaries, because the orbital energy and angular momentum are assumed to be conserved. Dissipation occurs because GWs carry energy and angular momentum away from the system, causing the binary to inspiral and eventually merge. For eccentric binaries, the GW energy and angular momentum fluxes have been computed to full 3PN order~\cite{Arun:2007sg, Arun:2009mc}. With these fluxes at hand and assuming small eccentricities ($e \lesssim 0.1$), Refs.~\cite{Yunes:2009yz,Tessmer:2010ii,Tessmer:2010sh,Moore:2016qxz} constructed time-domain and frequency domain waveforms to 2PN order. Waveform templates also exist for higher eccentricity systems ($ e \lesssim 0.4$), for example through the hybrid time-domain x-model of~\cite{Hinder:2008kv} and the hybrid frequency-domain model of~\cite{Huerta:2014eca}. These models, however are really not applicable to highly eccentric binaries.

The burst model previously developed~\cite{Loutrel:2014vja} is currently the only purely analytic model for highly elliptic orbits. We say the orbits are highly elliptic, and not highly eccentric because the latter implies that the eccentricity could be large and potentially greater than unity. On the other hand, ``highly elliptic'' indicates that we are always considering bound orbits. The burst model has currently only been developed to Newtonian order, which is an artifact of our desire to simplify our previous analysis as much as possible to be able to consider tests of GR. 

Nature, however, is not Newtonian, and thus, extending the burst model into the PN formalism serves two purposes: to improve its potential of aiding in the detection of highly elliptic binaries and to enhance its ability to perform interesting and important science. The burst model was conceived with the idea of testing GR. However, if there is an inherent modeling error within the GR model, it is possible that such modeling errors could fool us into believing we have detected a non-GR signal if we are not careful~\cite{Sampson:2013jpa, Yunes:2009ke}. Furthermore, the power stacking method is not immune to modeling error in the detection of signals and the extraction of their parameters~\cite{Tai:2014bfa}. Hence, the detection of such binaries and extraction of important astrophysics hinges on having an accurate prior to predict where the bursts occur in time-frequency space.

\subsection{Executive Summary}

We here extend the burst model developed in~\cite{Loutrel:2014vja} to higher PN order. The Newtonian burst model in~\cite{Loutrel:2014vja} focused on the bursts emitted during the inspiral of the binary only. Similarly, we here also focus on the inspiral of highly elliptic binaries within the PN framework. The motivation for developing a purely analytic model of the inspiral is the potential for the later construction of phenomenological inspiral-merger-ringdown models, a quasi-circular version of which played a pivotal role in the first gravitational wave observations by aLIGO~\cite{Abbott:2016blz}. Here, we treat the bursts as two dimensional regions of excess power in a time-frequency decomposition of a detectors data stream. We treat the bursts as boxes with characteristic time and frequency widths, which allows for a discretization of the time-frequency decomposition into tiles, with the burst being those tiles that contain excess power~\cite{Abbott:2016blz,Tai:2014bfa}. As with the Newtonian burst model, we characterize the sequence of bursts using the time and frequency centroids of the bursts, as well as the widths of the burst tiles, or alternatively the volume of the tiles, used to capture a certain amount of power within each bursts. These time-frequency observables are supplemented by a model describing the orbital evolution of the binary as a set of discrete, osculating Keplerian ellipses.

Similar to how the parameterized post-Einsteinian (ppE) burst sequence of~\cite{Loutrel:2014vja} was a parameterized deformation of a simplified GR sequence, the PN burst sequence will be a parameterized deformation of the Newtonian order sequence. The deformations will scale with an increasing power of a particular PN expansion parameter, which we choose to be the pericenter velocity $v_{p}$. The coefficients of a $k/2$-PN order term, which scales as $v_{p}^{k}$, are then a set of functions $[{\cal{V}}_{k}, {\cal{D}}_{k}, P_{k}, R_{k}]$ which are dependent on the physical parameters of the compact binary system. These functions correspond to the PN corrections to the rates of change of pericenter velocity and time eccentricity, and the expressions for the orbital period and pericenter distance, respectively. In this work, we neglect the spin of the compact objects and work in a point particle limit, such that these functions are only dependent on the time eccentricity $e_{t}$ and the symmetric mass ratio $\eta$. Hence, when working to $k/2$-PN order, one needs $4k$ functions $[{\cal{V}}_{k}(e_{t}, \eta), {\cal{D}}_{k}(e_{t}, \eta), P_{k}(e_{t}, \eta), R_{k}(e_{t}, \eta)]$ to parametrize all of the PN defomations.

We parametrize the PN burst sequence in time-frequency space by
\allowdisplaybreaks[4]
\begin{align}
\label{t-PN}
\frac{(t_{i} - t_{i-1})_{{\rm PN}}}{(t_{i} - t_{i-1})_{\rm N}} &= 1 + \vec{P}(e_{t,i}, \eta; v_{p}) \cdot \vec{X}(v_{p,i})
\\
\frac{f_{i}^{{\rm PN}}}{f_{i}^{\rm N}} &= 1 + \vec{R}^{(-1)}(e_{t,i}, \eta; v_{p}) \cdot \vec{X}(v_{p,i})
\\
\frac{\delta t_{i}^{{\rm PN}}}{\delta t_{i}^{k{\rm N}}} &= 1 + \vec{R}(e_{t,i}, \eta; v_{p}) \cdot \vec{X}(v_{p,i})
\\
\label{df-PN}
\frac{\delta f_{i}^{{\rm PN}}}{\delta f_{i}^{\rm N}} &= 1 + \vec{R}^{(-1)}(e_{t,i}, \eta; v_{p}) \cdot \vec{X}(v_{p,i})
\end{align} 
where $(t_{i}, f_{i})$ are the centroid of the bursts and $(\delta t_{i}, \delta f_{i})$ are the width and height of the tiles. We create the \emph{amplitude vector fields} $[\vec{P},\vec{R}]$, whose components are the functions $[P_{k}(e_{t}, \eta; v_{p}), R_{k}(e_{t}, \eta; v_{p})]$, which we further specify as implicit functions of the PN expansion parameter $v_{p}$ since their form changes depending on the choice of expansion parameter. The components of $\vec{R}^{(-1)}$ are defined such that
\begin{equation}
\left[1 + \vec{R}(e_{t}, \eta; v_{p}) \cdot \vec{X}(v_{p})\right]^{(-1)} \doteq 1 + \vec{R}^{(-1)}(e_{t}, \eta; v_{p}) \cdot \vec{X}(v_{p})\,,
\end{equation}
where the equality $\doteq$ should be understood as working in the limit of $v_{p} \ll 1$. The \emph{state vector} $\vec{X}(v_{p})$ contains the powers of $v_{p}$ that characterized each PN order corrections, specifically $\vec{X}(v_{p}) = (v_{p}, v_{p}^{2}, ..., v_{p}^{k})$. Hence the dot products provide the complete sum of all terms in a PN expansion up to $k/2$-PN order.

These time-frequency burst parameters, specifically $(t_{i}, f_{i}, \delta t_{i}, \delta f_{i})$, are functions of the symmetric mass ratio and the total mass of the binary, which are constant in time, as well as the pericenter velocity $v_{p,i}$ and eccentricity $e_{t,i}$ during each burst, which are evolving in time under the influence of radiation reaction. Hence, we must supplement the time-frequency sequence described above with the orbital evolution of the binary. To do this, we apply an osculating approximation that assumes the bursts are emitted instantaneously at pericenter, forcing the binary to move along a discrete set of Keplerian ellipses that osculate onto one another. The parameters of the $i$-th orbit will be functions of the parameters of the previous orbit, specifically
\begin{align}
\label{dvp-PN}
\frac{(v_{p,i} - v_{p,i-1})_{{\rm PN}}}{(v_{p,i} - v_{p,i-1})_{\rm N}} &= 1 + \vec{\cal{V}}(e_{t,i-1}, \eta; v_{p}) \cdot \vec{X}(v_{p,i-1})\,,
\\
\label{det-PN}
\frac{(\delta e_{t,i} - \delta e_{t,i-1})_{\rm PN}}{(\delta e_{t,i} - \delta e_{t,i-1})_{\rm N}} &= 1+ \vec{\cal{D}}(e_{t,i-1}, \eta; v_{p}) \cdot \vec{X}(v_{p,i-1})\,.
\end{align}
where we have introduced the two new amplitude vector fields $[\vec{\cal{V}}, \vec{\cal{D}}]$. In this work, we provide explicit expressions for the amplitude vector fields $\lambda_{\rm PN burst}^{a} \equiv (\vec{P}, \vec{R}, \vec{\cal{V}}, \vec{\cal{D}})$ complete to 3PN order. Equations~\eqref{t-PN}-\eqref{df-PN} and~\eqref{dvp-PN}-\eqref{det-PN} provide the complete PN burst model, which we use to calculate the burst model to 3PN order using the results for $\lambda_{\rm PN burst}^{a}$.

How does this new burst model aid us in the detection of highly elliptic binaries? In a realistic search, the burst model acts as a prior on where the bursts will occur in time-frequency space. For example, once a search detects a burst of power within an interferometer data stream (even if this burst of power is not ``loud'' enough to allow to claim detection), the burst model can then be used to search over "future" time-frequency space for successive bursts, as well as "past" time-frequency space for bursts that may have been missed by previous searches.  Physically, this amounts to searching over the parameters of the system that determine the prior, specifically the eccentricity and pericenter velocity during the initially detected burst, and the chirp mass and symmetric mass ratio of the binary.

The structure of the PN burst model should not be surprising given the structure of the ppE burst model in~\cite{Loutrel:2014vja}. The ppE model requires four exponent parameters $a_{i}$, which govern the power of $v_{p}$ of the corrections, and four amplitude parameters $\alpha_{i}$, which depend on the coupling constants of the theory and the eccentricity of the binary. In the PN formalism, the exponent parameter $k$ becomes a known quantity and only changes when one goes to higher order in the expansion variable. The amplitude parameters have now been replaced with four amplitude vector fields\footnote{These are not true vector fields, but are a set of scalar functions that have been combined into a discrete sequence. The terminology used for these functions goes along with the notation we have used to simplify some of the expressions in this work.} that parametrize the eccentricity and mass dependence of specific PN terms. As a result of this, rather than needing eight parameters as was the case in the ppE model, we require 4$k$ functions when working to $k/2$-PN order. These amplitude functions only depend on the initial eccentricity and the symmetric mass ratio, which together with the initial pericenter velocity and the chirp mass of the binary are the only parameters needed to define the model. Further, the fact that we require four vector fields to describe the burst model is a result of the fact that the model in only parametrized by four quantities: the orbital energy and angular momentum, and the energy and angular momentum fluxes of the GWs emitted by the system. Alternatively, as we will show, a different set of four parameters can be used: the orbital period, the mapping between the pericenter distance and pericenter velocity, and the rates of change of eccentricity and pericenter velocity due to radiation reaction. Working with these four quantities significantly improves the ease with which burst models can be constructed.

The remainder of this paper is dedicated to deriving the results presented above. In Section~\ref{review}, we review the Newtonian order burst model and present a simplified formalism used for the construction of the PN burst model. Section~\ref{PN} is dedicated to constructing a burst model at arbitrary PN order, which we later specialize to the cases of 1PN, 2PN, and 3PN orders. Section~\ref{discussion} discusses the results of the paper and their importance for future research. In this paper, we use geometric units where $G=c=1$.
\section{Constructing Burst Models}
\label{review}

This section is dedicated to reviewing how to create a burst model and the elements that go into such a model. We begin by reviewing the Newtonian burst model and how it was constructed. We then describe a new method of constructing burst models in general without any assumptions of the regime or theory of gravity we are working in. This new method greatly simplifies the construction of burst models, and will allow us to develop a completely generic GR PN model in the next section.
\subsection{The Newtonian Burst Model}
\label{sec:NewtBurstMod}

How do we construct a burst model? Recall that in Section~\ref{intro} we defined a burst model as a theoretical model prior to describe how the bursts evolve in time-frequency space. Such a model would tell us how the centroid and the volume of the bursts evolve in time and frequency from one burst to the next. But this evolution depends on the orbital parameters of the system, which themselves are also changing in time due to dissipative effects, such as the emission of GWs. Therefore, a complete burst model must provide a one-to-one mapping between the evolution of the system's physical parameters and how the bursts evolve in time and frequency. This requires the following ingredients:
\begin{enumerate}
\item {\textit{Orbital Evolution:}} A mapping that prescribes the evolution of the orbital parameters from one orbit to the next, including GW radiation-reaction.
\item {\textit{Centroid Mapping:}} A mapping that provides the time-frequency centroid of the burst $(t_{i}, f_{i})$, given the centroid of the previous burst $(t_{i-1}, f_{i-1})$, in terms of the orbital parameters of the system. 
\item{\textit{Volume Mapping:}} A mapping that describes how the time-frequency volume of the bursts changes from one to the next, in terms of the orbital parameters of the system. 
\end{enumerate}

\subsubsection{Orbit Evolution}

Let us start by reviewing how the ingredients listed above can be computed to leading (i.e.~Newtonian) order, focusing first on ingredient I (the orbital evolution). In Newtonian gravity, the orbital motion of two test particles can be described though Keplerian ellipses, which are characterized by two conserved quantities, the orbital energy $E$ and the orbital angular momentum $L$. Alternatively, one can parameterize any such orbit in terms of its pericenter distance $r_{p}$ and its orbital eccentricity $e$, which are related to $E$ and $L$ at Newtonian order in a PN expansion by
\allowdisplaybreaks[4]
\begin{align}
\label{EN}
E &= - \frac{M^{2} \eta \left(1 - e\right)}{2 r_{p}}\,,
\\
\label{LN}
L &= \eta \sqrt{M^{3} r_{p} \left(1 + e\right)}\,.
\end{align}
where $\eta = m_{1} m_{2}/M^{2}$ is the symmetric mass ratio and $M = m_{1}+m_{2}$ is the total mass of the system. 

The burst model requires knowledge of how $(E,L)$ or $(r_{p},e)$ evolve from one orbit to the next. Due to the nature of the emission of gravitational radiation in highly elliptic systems, we may treat the problem as a set of Keplerian orbits that change effectively instantaneously at pericenter, allowing the orbits to \emph{osculate} onto one another. Hence, we may write
\begin{align}
E_{i} &= E_{i-1} + \Delta E_{(i,i-1)}\,,
\\
L_{i} &= L_{i-1} + \Delta L_{(i,i-1)}\,,
\end{align}
where $\Delta E_{(i,i-1)}$ and $\Delta L_{(i,i-1)}$ are the changes in orbital energy and angular momentum due to GW emission from one orbit to the next, and the labels represent which orbit the above quantities are evaluated on. By ``osculating orbits,'' we mean that the elements of the Keplerian orbit are constant throughout the orbit except at pericenter, where they change drastically and the new elements define a new Keplerian orbit. In this approximation, one thus treats the radiation, and all changes generated by it, as arising instantaneously at pericenter.

In general, the total change in energy and angular momentum between times $T_{i-1}$ and $T_{i}$ due to GW emission is given by
\begin{align}
\label{change-E}
\Delta E_{(i,i-1)} &= \int_{T_{i-1}}^{T_{i}} \dot{E}\left(r_{p}, e, \psi\right) dt\,,
\\
\label{change-L}
\Delta L_{(i,i-1)} &= \int_{T_{i-1}}^{T_{i}} \dot{L}\left(r_{p}, e, \psi\right) dt\,,
\end{align}
where $\psi$ is the true anomaly, $T_{i-1}$ and $T_{i}$ are the times of consecutive pericenter passages, and the dot refers to derivatives with respect to time. At Newtonian order, the GW energy and angular momentum fluxes, $\dot{E}$ and $\dot{L}$, are given, for example, by Eq.~(12.78) in~\cite{PW}. The fluxes are functions of the orbital elements, which are themselves functions of time through the true anomaly $\psi$. Thus, the above definitions would have to be supplemented with the time evolution of $\psi$ itself, namely $\dot{\psi}\left(r_{p}, e, \psi\right)$. In addition, since the fluxes depend on the true anomaly, they contain gauge-dependent terms~\cite{PW}. However, these terms vanish upon integration, leaving $\Delta E_{(i,i-1)}$ and $\Delta L_{(i,i-1)}$ independent of the radiation reaction gauge.

To evaluate the integrals in Eqs.~\eqref{change-E} and~\eqref{change-L}, we perform a change of variable from $t$ to the true anomaly, using  $dt = d\psi / \dot{\psi}$. The new limits of integration become $\psi_{i-1}$ and $\psi_{i} = \psi_{i-1} + 2 \pi$, or more simply $[0,2\pi].$ The orbital elements now depend on the true anomaly rather than time, which simplifies the integrands, but this is still not enough to evaluate them analytically. To do so, we use the fact that the orbits are osculating and the GW emission occurs instantaneously at pericenter, which ensures that $r_{p}$ and $e$ are constant everywhere except at closest approach. With this, the integrals become
\begin{align}
\label{eq:DeltaE-int}
\Delta E_{(i,i-1)} &= \int_{0}^{2\pi} \frac{\dot{E}\left(r_{p,i-1}, e_{i-1}, \psi\right)}{\dot{\psi}\left(r_{p,i-1}, e_{i-1}, \psi\right)} d\psi
\\
\label{eq:DeltaL-int}
\Delta L_{(i,i-1)} &= \int_{0}^{2\pi} \frac{\dot{L}\left(r_{p,i-1}, e_{i-1}, \psi\right)}{\dot{\psi}\left(r_{p,i-1}, e_{i-1}, \psi\right)} d\psi
\end{align}
which can be evaluated analytically.

Alternatively, we can exploit the definition of orbital averaged GW fluxes to rewrite these changes in a simpler way. The orbital averaged energy flux, for example, is given by $\langle \dot{E} \rangle \equiv \Delta E_{(i,i-1)} / P$, where $P$ is the orbital period of the binary, and likewise for the angular momentum flux. With this definition, we are free to write
\begin{align}
\label{E-next}
E_{i} &= E_{i-1} + P_{i-1} \langle \dot{E} \rangle_{i-1}\,,
\\
\label{L-next}
L_{i} &= L_{i-1} + P_{i-1} \langle \dot{L} \rangle_{i-1}\,.
\end{align}
Indeed, we recognize the integral expressions in Eqs.~\eqref{eq:DeltaE-int} and~\eqref{eq:DeltaL-int} as simply the product of the orbital period and $\langle \dot{E} \rangle_{i-1}$ or $\langle \dot{L} \rangle_{i-1}$ by definition. It might seem odd that orbit averaged quantities appear in the above expressions since the GW emission is happening mostly during pericenter passage, and thus smearing the emission over the entire orbit would appear incorrect. However, this is purely a result of the \emph{definition} of the orbital averaged quantities, and has nothing to do with the nature of the GW emission or the validity of the orbital averaged approximation for the systems we are considering~\cite{Loutrel-avg}.

The orbital energy and angular momentum have a clear physical meaning, but $r_{p}$ and $e$ allow us to straightforwardly visualize the geometry of the system that is generating the bursts (at least at Newtonian order). At this order, it does not matter which set of quantities, $(E,L)$ or $(r_{p}, e)$, we decide to use for the orbital evolution. For the Newtonian model in~\cite{Loutrel:2014vja}, we decided to use $(r_{p}, e)$, so let us continue to do so here. We need to solve the system given by Eqs.~\eqref{EN} and~\eqref{LN} for the functionals $r_{p}\left(E, L\right)$ and $e\left(E, L\right)$. To obtain the evolution of the pericenter distance and the orbital eccentricity, we evaluate the functionals at the desired orbit, specifically $r_{p, i} = r_{p} \left[E_{i}\left(E_{i-1}, L_{i-1}\right), L_{i}\left(E_{i-1}, L_{i-1}\right)\right]$ and $e_{i} = e\left[E_{i}\left(E_{i-1}, L_{i-1}\right), L_{i}\left(E_{i-1}, L_{i-1}\right)\right]$. Evaluating the functionals with Eqs.~\eqref{E-next} and~\eqref{L-next} gives
\begin{align}
\label{burst-rp-N}
r_{p,i} &= r_{p,i-1} \left[1 - \frac{59 \pi \sqrt{2}}{24} \eta \left(\frac{M}{r_{p,i-1}}\right)^{5/2} \left(1 + \frac{121}{236} \delta e_{i-1}\right)\right]\,,
\\
\label{burst-e-N}
\delta e_{i} &= \delta e_{i-1} + \frac{85 \pi \sqrt{2}}{12} \eta \left(\frac{M}{r_{p,i-1}}\right)^{5/2} \left(1 - \frac{1718}{1800} \delta e_{i-1}\right)\,,
\end{align}
where we have kept only leading-order terms in $\delta e \equiv 1 - e \ll 1$ and in $M/r_{p} \ll 1$ in all expressions (since we are working to Newtonian order). These equations recursively describe how the orbit shrinks and circularizes as the binary inspirals, thus completely describing the orbital evolution.
\subsubsection{Centroid Mapping}
The second ingredient we need for any burst model is the centroid mapping. The centroids of the bursts are given by the set $(t_{i}, f_{i})$ at which the bursts occur. What we desire is the mapping $t_{i-1} \rightarrow t_{i}$ and $f_{i-1} \rightarrow f_{i}$. Since the orbits are osculating, the time between bursts is trivially given by the orbital period $P$ to Newtonian order, which is given by
\begin{equation}
\label{Torb}
P = \frac{2 \pi r_{p}^{3/2}}{M^{1/2} (1 - e)^{3/2}}\,.
\end{equation}
 Thus to obtain the time mapping, we simply have to evaluate the orbital period at the desired orbit,
\begin{equation}
t_{i} = t_{i-1} + \frac{2 \pi}{M^{1/2}} \left[\frac{r_{p,i} \left(r_{p,i-1}, \delta e_{i-1}\right)}{\delta e_{i}\left(r_{p,i-1}, \delta e_{i-1}\right)}\right]^{3/2}
\end{equation}
where the mappings $r_{p,i}\left(r_{p,i-1}, \delta e_{i-1}\right)$ and $\delta e_{i}\left(r_{p,i-1}, \delta e_{i-1}\right)$ are given by Eqs.~\eqref{burst-rp-N} and~\eqref{burst-e-N}, respectively. 

The fact that the bursts are separated by an orbital period can be seen more generally by writing $\dot{\psi} = \dot{\psi}_{\text{cons}} + \dot{\psi}_{\text{diss}}$, where $\dot{\psi}_{\text{cons}}$ is the conservative part coming from Keplerian orbital dynamics, and $\dot{\psi}_{\text{diss}}$ is the dissipative piece that comes from radiation reaction. The time between successive pericenter passages, i.e. the orbital period, is then
\begin{equation}
\label{change-t}
t_{i} - t_{i-1} = \int_{0}^{2\pi} \frac{d\psi}{\dot{\psi}}\,.
\end{equation}
The dissipative part contains terms that depend on the radiation reaction gauge, which vanish upon integration. Thus we are left with only the conservative piece of $\dot{\psi}$, and by assuming the orbits are osculating and the GW emission is instantaneous, this evaluates upon integration to the orbital period for an unperturbed, purely conservative orbit. This does not mean that the orbital period is not evolving. GW emission carries energy and angular momentum away from the binary that changes the orbital period. The dissipative part of the above integral does vanish, but the dissipative part of $P$, namely $\dot{P}$, does not. The above result simply implies that the time between pericenter passages is the orbital period of a Keplerian orbit, since radiation reaction is happening rapidly around pericenter.

The GW frequency on the other hand requires knowledge of the Fourier transform of the GWs emitted during each burst, the \emph{Fourier-domain waveform}. From Fig.~7 in~\cite{Turner:1977}, the GW power is highly peaked around
\begin{align}
\label{f-GW}
f_{\rm GW} = \frac{1}{2 \pi \tau_{\rm GW}}\,,
\end{align}
where $\tau_{\rm GW}$ is the characteristic GW time, defined by
\begin{align}
\label{tau-GW}
\tau_{\rm GW} &\equiv \frac{\text{pericenter distance}}{\text{pericenter velocity}}\,.
\end{align}
At Newtonian order, the pericenter velocity is given by
\begin{equation}
\label{vp-N}
v_{p} = \sqrt{\frac{M (1+e)}{r_{p}}}
\end{equation}
and thus $\tau_{\rm GW}$ is
\begin{align}
\tau_{\rm GW} &= \frac{r_{p}^{3/2}}{\left[M (1 + e)\right]^{1/2}}\,,
\end{align}
which roughly corresponds to the amount of time the system spends at pericenter. This time is a functional of the pericenter distance and eccentricity; hence, to obtain the frequency of the $i$-th burst $f_{i}$, one simply has to evaluate the characteristic GW time associated with the orbit $(r_{p,i}, e_{i})$, where the mapping to the parameters of the previous orbit are given by Eqs.~\eqref{burst-rp-N} and~\eqref{burst-e-N}:
\begin{equation}
f_{i} = \frac{M^{1/2} \left[2 - \delta e_{i}\left(r_{p,i-1}, \delta e_{i-1}\right)\right]^{1/2}}{2 \pi \left[r_{p,i}\left(r_{p,i-1}, \delta e_{i-1}\right)\right]^{3/2}}\,.
\end{equation}
This completes the mapping of the time-frequency centroid of the bursts. 

The prescription we provide for the frequency of the bursts is dependent on the frequency domain waveform, or alternatively the GW power, peaking at $\tau_{\rm GW}^{-1}$. This intuition comes from~\cite{Turner:1977}, where for parabolic orbits, and at Newtonian order, it is shown that the GW power peaks roughly at $\tau_{\rm GW}^{-1}$. However, for circular binaries, the power peaks at twice the orbital frequency, and it can easily be checked that the prescription given above does not reproduce this result when $e=0$. This implies that there are uncontrolled remainders that depend on $\delta e$ that correct the above expression to account for this. However, because we are working in the limit where $\delta e \ll 1$, we expect such corrections to be subdominant.

\subsubsection{Volume Mapping}

The last ingredient we need is the volume mapping. The bursts are not instantaneously emitted at pericenter and are not solely peaked at one frequency. The emission is instead spread out over the full pericenter passage and over multiple frequencies. To complete the burst model, we need to determine how the time-frequency size of the bursts change from one to another. We may describe the bursts as any two dimensional objects in time and frequency. For simplicity, we chose to model the bursts as boxes with widths
\begin{align}
\label{delta-t}
\delta t = \xi_{t} \tau_{\rm GW}
\\
\label{delta-f}
\delta f= \xi_{f} f_{\rm GW}
\end{align}
where $\xi_{t}$ and $\xi_{f}$ are constants of proportionality that are chosen from data analysis considerations. For example, one can choose these constants such that a desired percentage of the GW power ($90\%$ for example) is contained in each box. More general two dimensional objects, such as ellipsoids, could be used for this construction, but boxes are the simplest. To obtain how these widths change from one burst to the next, we simply have to evaluate Eqs.~\eqref{delta-t}-\eqref{delta-f} at the parameters of the orbit $(r_{p,i}, e_{i})$: 
\begin{align}
\label{eq:delta-t-eval}
\delta t_{i} &= \frac{\xi_{t} \left[r_{p,i}\left(r_{p,i-1}, \delta e_{i-1}\right)\right]^{3/2}}{M^{1/2} \left[2 - \delta e_{i-1} \left(r_{p,i-1}, \delta e_{i-1}\right)\right]^{1/2}}
\\
\label{eq:delta-f-eval}
\delta f_{i} &= \frac{\xi_{f} M^{1/2} \left[2 - \delta e_{i}\left(r_{p,i-1}, \delta e_{i-1}\right)\right]^{1/2}}{2 \pi \left[r_{p,i}\left(r_{p,i-1}, \delta e_{i-1}\right)\right]^{3/2}}\,.
\end{align}
Note that in the case of ellipsoids, the results are the same, but these quantities can instead be interpreted as the semi-minor and semi-major axes of the ellipsoids. For a realistic search, ellipsoids would actually be more appropriate choice since they are a more accurate representation of the time-frequency structure of the bursts. However, for the purposes of this work, this choice is irrelevant, as the goal is to characterize the two dimensional objects via the scales in Eqs.~\eqref{eq:delta-t-eval} and~\eqref{eq:delta-f-eval}. This completes the review of the construction of the burst model to Newtonian order. 

\subsection{A Simplified Formalism}
\label{simp}

Ultimately, we are interested in a PN burst model at generic (presumably very high) PN order. Building a generic order PN model by following the construction above might at first seem like an intractable problem. To start, one would have to take the orbital energy $E_{\rm PN}$ and angular momentum $L_{\rm PN}$ at arbitrary order and invert these expressions to obtain $r_{p}(E_{\rm PN}, L_{\rm PN})$ and $e(E_{\rm PN}, L_{\rm PN})$. Then, one would need to use the energy and angular momentum fluxes to compute the evolution of the orbital energy and angular momentum. From there, the pericenter and orbital eccentricity mapping would have to be computed using the functionals $r_{p,i}\left[E_{i}(E_{i-1}, L_{i-1}), L_{i}(E_{i-1}, L_{i-1})\right]$ and $e_{i}\left[E_{i}(E_{i-1}, L_{i-1}), L_{i}(E_{i-1}, L_{i-1})\right]$. While this may actually be possible from a mathematical standpoint, it will be very non-trivial to do so at arbitrary order. Thus, in this subsection, we will instead seek a simplified formalism that is more practical to implement.

The new method we seek must be more direct than the previous method discussed, removing steps that are redundant and reducing the number of physical quantities we need to work with. We begin by noting a number of assumptions that we will use to simplify the analysis:
\begin{enumerate}
\item {\textit{Osculating Orbits:}} Any changes in the orbital parameters will be modeled as occurring instantaneously around pericenter passage, leaving the orbital parameters constant throughout the rest of the orbit.
\item {\textit{High Ellipticity:}} The orbits we consider are highly elliptical, so we define a small parameter $\delta e_{t} \equiv 1 - e_{t}$ and work perturbatively in the regime $\delta e_{t} \ll 1$.
\item{\textit{PN Orbits:}} We will work within the PN framework, expanding all expressions in the pericenter velocity $v_{p} \ll 1$.
\end{enumerate}

The first and second assumptions follow directly from the nature of the systems we consider in this paper. Note that in the second assumption we are now working with the time eccentricity from the QK parametrization. The reason for this is that in PN theory, there is no unique concept for the orbital eccentricity, as there are actually three eccentricities that enter the QK equations of motion, specifically $(e_{t}, e_{r}, e_{\phi})$. All three of these eccentricities reduce to the orbital eccentricity in the Newtonian limit, but they are distinct quantities within PN theory. We choose to work with $e_{t}$ and will express all quantities in terms of it. We discuss this in more detail in~\ref{fields}.

The final assumption is new to this analysis and replaces the previous Newtonian assumption. At Newtonian order, we worked with the pericenter distance $r_{p}$ as one of our physical parameters. We will now choose to work with the pericenter velocity $v_{p}$ instead. This change is meant to put the computation more in line with the standard PN formalism for quasi-circular inspirals, as well as to remove some difficulties that result in there being terms that depend on half-integer powers of $r_{p}$ in the dissipative sector. The mapping between the pericenter distance and velocity is given explicitly to 1PN order in Eqs.~\eqref{rp-1PN} and~\eqref{R2}, with the 2PN and 3PN corrections given in Eqs.~\eqref{R4} and~\eqref{R6}. With the above assumptions, any burst model requires the three ingredients laid out in Sec.~\ref{sec:NewtBurstMod}. 

Let's start with the orbital evolution, where now we focus on the evolution of the pericenter velocity and the orbital eccentricity. Rather than starting from the orbital energy and angular momentum, we are free to write the velocity and eccentricity mappings as
\begin{align}
\label{vp-map}
v_{p,i} &= v_{p,i-1} + \Delta v_{p, (i,i-1)}\,,
\\
\label{e-map}
\delta e_{t,i} &= \delta e_{t,i-1} - \Delta e_{t,(i,i-1)}\,,
\end{align}
where $\Delta v_{p, (i,i-1)}$ and $\Delta e_{t,(i,i-1)}$ are the change in pericenter velocity and time eccentricity between two successive orbits, and we have used the fact that $\delta e_{t} = 1 - e_{t}$ to write $\Delta \delta e_{t} = - \Delta e_{t}$. These mappings are directly analogous to the mappings of energy and angular momentum in our Newtonian model, given by Eqs.~\eqref{change-E} and~\eqref{change-L}.

Expressions for $\Delta v_{p, (i,i-1)}$ and $\Delta e_{t,(i,i-1)}$ can be found in exactly the same way as in Eqs.~\eqref{eq:DeltaE-int} and~\eqref{eq:DeltaL-int}. We may thus jump ahead and directly write
\begin{align}
\Delta v_{p,(i,i-1)} &= \int_{0}^{2\pi} \frac{\dot{v}_{p}\left(v_{p,i-1}, e_{t,i-1}, \psi\right)}{\dot{\psi}\left(v_{p,i-1}, e_{t,i-1}, \psi\right)} d\psi
\\
\Delta e_{t,(i,i-1)} &= \int_{0}^{2\pi} \frac{\dot{e}_{t}\left(v_{p,i-1}, e_{t,i-1}, \psi\right)}{\dot{\psi}\left(v_{p,i-1}, e_{t,i-1}, \psi\right)} d\psi
\end{align}
where $\dot{v}_{p}$ and $\dot{e}$ are the rates of change of pericenter velocity and orbital eccentricity. These rates, once again, depend on the true anomaly and thus have gauge-dependent terms arising from the GW sector. Upon integration, these terms vanish, except now the above quantities are not necessarily gauge-invariant as they depend on the specific coordinate system one chooses to do the PN calculation in.

To our knowledge, the expressions $\dot{v}_{p}\left(v_{p}, e_{t}, \psi\right)$ and $\dot{e}_{t}\left(v_{p}, e_{t}, \psi\right)$ have not yet been explicitly computed and would not be easy to compute, which would leave something of a gap in constructing the orbit evolution for our bursts. However, we may once again exploit the definition of orbit averaging and write Eqs.~\eqref{vp-map} and~\eqref{e-map} as
\begin{align}
\label{vp-burst}
v_{p,i} &= v_{p,i-1} + P_{i-1} \langle \dot{v}_{p} \rangle_{i-1}\,,
\\
\label{e-burst}
\delta e_{t,i} &= \delta e_{t,i-1} - P_{i-1} \langle \dot{e}_{t} \rangle_{i-1}\,.
\end{align}
The orbit averaged quantities $\langle \dot{v}_{p} \rangle$ and $\langle \dot{e}_{t} \rangle$ can be easily computed from the orbital energy and angular momentum and the corresponding fluxes, which are known to full 3PN order. We have thus completely constructed the orbit evolution for our burst model.

We now focus on the centroid and volume mappings. Once again, we will treat the bursts as boxes in time and frequency, and determine the mapping between the centroids and widths of the boxes. The characteristic GW time is still given by Eq.~\eqref{tau-GW}. In the Newtonian model, we used the expression $v_{p}\left(r_{p}, e_{t}\right)$ given by Eq.~\eqref{vp-N} to write this time in terms of $(r_{p}, e_{t})$. Since we are now working with $v_{p}$ instead of $r_{p}$, we can invert the relationship between these two parameters to obtain $r_{p}\left(v_{p}, e_{t}\right)$,which is given explicitly in Eqs.~\eqref{rp-1PN}-\eqref{R2} and~\eqref{R4}-\eqref{R6}, and write $\tau_{\rm GW}$ in terms of $(v_{p}, e_{t})$. Once this time is specified, we may define the characteristic GW frequency by Eq.~\eqref{f-GW}. The centroid and volume mappings follow the exact same analysis as the Newtonian model, only parameterized by the pericenter velocity rather than the pericenter distance. Hence, we may write
\begin{align}
\label{change-t}
t_{i} &= t_{i-1} + P\left[v_{p,i}\left(v_{p,i-1}, e_{t,i-1}\right), e_{t,i} \left(v_{p,i-1}, e_{t,i-1}\right)\right],
\\
\label{change-f}
f_{i} &= f_{\rm GW} \left[v_{p,i}\left(v_{p,i-1}, e_{t,i-1}\right), e_{t,i}\left(v_{p,i-1}, e_{t,i-1}\right)\right],
\\
\label{change-dt}
\delta t_{i} &= \delta t \left[v_{p,i}\left(v_{p,i-1}, e_{t,i-1}\right), e_{t,i}\left(v_{p,i-1}, e_{t,i-1}\right)\right],
\\
\label{change-df}
\delta f_{i} &= \delta f \left[v_{p,i}\left(v_{p,i-1}, e_{t,i-1}\right), e_{t,i}\left(v_{p,i-1}, e_{t,i-1}\right)\right],
\end{align}
thus completing the last two ingredients we need for our simplified formalism.

\section{A Generic PN Formalism}
\label{PN}
With the application of assumption I, we have constructed a purely generic burst model through Eqs.~\eqref{vp-burst},~\eqref{e-burst}, and~\eqref{change-t}-\eqref{change-df} that applies in any theory of gravity. We now seek to use this formalism to create a burst model at generic PN order. We will provide explicit expressions for the burst model at 1PN, 2PN, and 3PN orders in Sec.~\ref{examples}.

The above considerations imply that, to construct our burst model, we need PN expansions for four quantities: the orbital period, the pericenter distance, the rate of change of pericenter velocity, and the rate of change of orbital eccentricity. We can write these expansions to arbitrary PN order as
%
%
\begin{align}
\label{Torb-PN}
P^{\rm PN} &= P^{\rm N}\left(v_{p}, e_{t}\right) \left[1 + \vec{P}(e_{t}, \eta; v_{p}) \cdot \vec{X}(v_{p})\right]\,,
\\
\label{vp-PN}
r_{p}^{\rm PN} &= r_{p}^{\rm N}\left(v_{p}, e_{t}\right) \left[1 + \vec{R}(e_{t}, \eta; v_{p}) \cdot \vec{X}(v_{p})\right]\,,
\\
\label{vpdot-PN}
\langle \dot{v}_{p}^{\rm PN} \rangle &= \langle \dot{v}_{p}^{\rm N} \rangle \left(v_{p}, e_{t}\right) \left[1 + \vec{V}(e_{t}, \eta; v_{p}) \cdot \vec{X}(v_{p})\right]\,,
\\
\label{edot-PN}
\langle \dot{e}_{t}^{\rm PN} \rangle &= \langle \dot{e}_{t}^{\rm N} \rangle \left(v_{p}, e_{t}\right) \left[1 + \vec{E}(e_{t}, \eta; v_{p}) \cdot \vec{X}(v_{p})\right]\,,
\end{align}
with the Newtonian order quantities
\begin{align}
\label{Porb-N}
P^{\rm N} &= \frac{2 \pi M}{v_{p}^{3}} \left(\frac{1+e_{t}}{1-e_{t}}\right)^{3/2}\,,
\\
\label{rp-N}
r_{p}^{\rm N} &= \frac{M (1+e_{t})}{v_{p}^{2}}\,,
\\
\label{vp-dot-N}
\langle \dot{v}_{p}^{\rm N} \rangle &= \frac{32}{5} \frac{\eta}{M} v_{p}^{9} \frac{\left(1 - e_{t}\right)^{3/2}}{\left(1 + e_{t}\right)^{15/2}} V_{\rm N}(e_{t})\,,
\\
\label{e-dot-N}
\langle \dot{e}_{t}^{\rm N} \rangle &= -\frac{304}{15} e_{t} \frac{\eta}{M} v_{p}^{8} \frac{\left(1 - e_{t}\right)^{3/2}}{\left(1 + e_{t}\right)^{13/2}} \left(1 + \frac{121}{304} e_{t}^{2}\right)\,,
\\
V_{\rm N}(e) &= 1 - \frac{13}{6} e_{t} + \frac{7}{8} e_{t}^{2} - \frac{37}{96} e_{t}^{3}\,.
\end{align}

We refer to the vector $\vec{X}$ as the PN \emph{state vector}, which depends on the PN expansion parameter. In our case, the PN expansion parameter is $v_{p}$ and the components of $\vec{X}$ are simply $X_{k} = v_{p}^{k}$. Furthermore, we refer to the vector fields $(\vec{P}, \vec{R}, \vec{V}, \vec{E})$ as PN \emph{amplitude vectors}, which are functions of the orbital eccentricity and the symmetric mass ratio. In Eq.~\eqref{Torb-PN}-\eqref{edot-PN}, we have chosen to include a $v_{p}$ label in the amplitude vectors to remind us that the functional form of its components depends on the parameter one expands about, i.e.~if we had chosen to work with the $x$ PN expansion parameter instead of $v_{p}$, then the eccentricity and symmetric mass ratio dependence of the PN amplitude vectors would be different. The dot products between the state vectors and the amplitude vectors take the simple form
\begin{equation}
\vec{A}(e_{t}, \eta; v_{p}) \cdot \vec{X}(v_{p}) = \sum_{k=2}^{\infty} A_{k}(e_{t}, \eta; v_{p}) v_{p}^{k}
\end{equation}
where $\vec{A} \in (\vec{P}, \vec{R}, \vec{V}, \vec{E})$. We recognize the above expression as the summation of PN corrections to the associated quantity. The summation index $k$ acts as the PN order of each term and starts at $k=2$ corresponding to the corrections at 1PN order. The components of $\vec{A}$, specifically $A_{k}(e_{t},\eta)$, we then recognize as the coefficient of the $k/2$-PN order term.

The components of the amplitude vectors $(\vec{P}, \vec{R}, \vec{V}, \vec{E})$ can be easily computed from known PN quantities. As an example, consider the orbital period. This quantity can be written as a function of the reduced energy $\varepsilon$ and angular momentum $j$ through the equation $P=2 \pi / n$, where $n$ is the mean motion, given to 3PN order by Eq.~(348a) in~\cite{Blanchet:2013haa}. In turn, the reduced energy and angular momentum can also be written in terms of the pericenter velocity and eccentricity, $\varepsilon(v_{p}, e_{t})$ and $j(v_{p}, e_{t})$, which can be inserted into the expression $P(\varepsilon, j)$ and expanded about $v_{p} \ll 1$. The coefficients of each power of $v_{p}$ are then the components of the vector field $P_{k}$. We will provide expressions for these components at specific PN orders when we construct burst models at specific PN orders.

This should be very reminiscent of the computations commonly carried out in the ppE formalism. In the latter,  four deformations characterized by eight parameters (4 \emph{amplitude parameters}, which are actually fields since they depend on the eccentricity of the binary, and 4 \emph{exponent parameters}) are used to completely specify the burst model in a general theory of gravity that reduces to GR in the low-velocity/weak-field limit. Instead of the four amplitude parameters of the ppE model, we now have four amplitude vectors fields $\lambda_{\rm PN}^{a} = (\vec{P}, \vec{R}, \vec{V}, \vec{E})$, which characterize the PN corrections to the Newtonian quantities. Also, the ppE exponent parameters have been replaced by the PN exponent $k$, which is a known number. Hence, instead of the eight ppE parameters, we need $4k$ PN functions when working to $k/2$-PN order. Each of the PN vector fields is a function of the parameters of the system, which we have written solely as functions of the eccentricity and the symmetric mass ratio. This will be true at 1PN order, but at higher PN order, the functions can depend on other physical parameters, such as the spins of compact objects, or the equation of state of supranuclear matter when at least one of the binary components is a NS.

The goal of this section will be to write the PN modifications to the Newtonian mappings in terms of the set of PN functions $\lambda_{{\rm PN},k}^{a} = (P_{k}, V_{k}, R_{k}, E_{k})$. We begin with the first ingredient, the orbital evolution, specified in our simplified formalism by Eqs.~\eqref{vp-burst} and~\eqref{e-burst}. In particular, we concentrate first on the evolution of the pericenter velocity. By exploiting the definition of orbit averaging, we are able to write the change in this quantity as $P \; \langle \dot{v}_{p} \rangle$, which is exactly the second term in Eq.~\eqref{vp-burst}. Hence, to obtain the velocity mapping, we simply have to multiply Eqs.~\eqref{Torb-PN} and~\eqref{vpdot-PN} together and expand in $v_{p}$. It is not difficult to see that our expansion is a product of two sums that is equivalent to a double sum of the form
\begin{align}
\left(\vec{P} \cdot \vec{X}\right) \left(\vec{V} \cdot \vec{X}\right) &= \left(\sum_{k=2}^{\infty} P_{k} \; v_{p}^{k}\right) \left(\sum_{k=2}^{\infty} V_{k} \; v_{p}^{k}\right) 
\nn \\
&= \sum_{k=2}^{\infty}  \sum_{j=2}^{k-2} 
P_{k-j} \; V_{j} \; v_{p}^{k}\,,
\nn \\
&= \left(\vec{P} \circ \vec{V}\right) \cdot \vec{X}\,,
\end{align}
where we have used the definition of the Cauchy product to rewrite the product of the sums as the discrete convolution of two series. When $k-2 < 2$, the convolution is exactly zero. Using this result, we write the velocity mapping as
\begin{align}
\label{vp-map-PN}
v_{p,i} &= v_{p,i-1} \left\{1 + \frac{\pi}{5} \eta v_{p, i-1}^{5} {\cal{V}}_{\rm N}\left(\delta e_{t,i-1}\right) \left[1 + \vec{{\cal{V}}}\left(\delta e_{t,i-1}, \eta; v_{p}\right) \cdot \vec{X}(v_{p,i-1})\right]\right\}
\end{align}
where the Newtonian term in this expression is
\begin{align} 
\label{VN}
V_{\rm N}(\delta e_{t}) &= - \frac{65}{96} + \frac{151}{96} \delta e_{t} - \frac{9}{32} \delta e_{t}^{2} + \frac{37}{96} \delta e_{t}^{3}\,,
\\
{\cal{V}}_{\rm N}(\delta e_{t}) &= \frac{V_{\rm N}(\delta e_{t})}{\left(1 - \frac{1}{2} \delta e_{t}\right)^{6}}
\nn \\
&= - \frac{65}{96} - \frac{11}{24} \delta e_{t} + {\cal{O}}\left(\delta e_{t}^{2}\right)
\end{align}
and the new amplitude vector $\vec{\cal{V}}$ is
\begin{align}
\label{calV}
\vec{{\cal{V}}}(\delta e_{t}, \eta; v_{p}) &= \vec{V}(\delta e_{t}, \eta; v_{p}) + \vec{P}(\delta e_{t}, \eta; v_{p}) + \vec{P}(\delta e_{t}, \eta; v_{p}) \circ \vec{V}(\delta e_{t}, \eta; v_{p})\,,
\end{align}
which should be expanded about $\delta e_{t} \ll 1$ by Assumption II. In the above expression, the pericenter velocity is decreasing from one orbit to the next for highly elliptic orbits at Newtonian order. We refer to this behavior as \textit{pericenter braking}, which will be explored in more detail in Section~\ref{brake}. 

We may follow the same procedure for the eccentricity mapping to find
\begin{align}
\label{e-map-PN}
\delta e_{t,i} &= \delta e_{t,i-1} + \frac{85 \pi}{48} \eta v_{p,i-1}^{5} {\cal{D}}_{\rm N}(\delta e_{t,i-1})  \left[1 + \vec{\cal{D}}(\delta e_{t,i-1}, \eta; v_{p}) \cdot \vec{X}(v_{p,i-1})\right]\,,
\end{align}
with the Newtonian function
\begin{align}
{\cal{D}}_{\rm N}(\delta e) &= \frac{(1 - \delta e_{t}) \left(1 - \frac{242}{425} \delta e_{t} + \frac{121}{425} \delta e_{t}^{2}\right)}{\left(1 - \frac{1}{2} \delta e_{t}\right)^{5}}\,,
\nn \\
&= 1 + \frac{791}{850} \delta e_{t} + {\cal{O}}\left(\delta e_{t}^{2}\right)\,,
\end{align}
and the amplitude vector 
\begin{align}
\label{calD}
\vec{{\cal{D}}}(\delta e_{t}, \eta; v_{p}) &= \vec{E}(\delta e_{t}, \eta; v_{p}) + \vec{P}(\delta e_{t}, \eta; v_{p}) + \vec{P}(\delta e_{t}, \eta; v_{p}) \circ \vec{E}(\delta e_{t}, \eta; v_{p})\,.
\end{align}
We thus find that the PN amplitude vectors $(\vec{\cal{V}}, \vec{\cal{D}})$ can be expressed in terms of the known PN amplitude vectors $(\vec{P}, \vec{V}, \vec{E})$. The above expressions are purely generic within the PN formalism, allowing them to be applied at any PN order.

Now, let us consider the second ingredient of the PN burst model: the centroid mapping. The GW time is given in Eq.~\eqref{tau-GW}, while the pericenter distance is given in Eq.~\eqref{vp-PN}. We thus have that the GW time at arbitrary PN order is
\begin{equation}
\tau_{\rm GW} = \frac{M \left(2 - \delta e_{t}\right)}{v_{p}^{3}} \left[1 + \vec{R}(\delta e_{t}, \eta; v_{p}) \cdot \vec{X}(v_{p})\right]\,,
\end{equation}
and the frequency mapping between boxes is
\begin{align}
\label{f-map-PN}
f_{i} &= \frac{\left[v_{p,i} \left(v_{p,i-1}, \delta e_{t,i-1}\right)\right]^{3}}{2 \pi M \left[2 - \delta e_{t,i} \left(v_{p,i-1}, \delta e_{t,i-1}\right)\right]} \left\{1 + \vec{R}^{(-1)}\left[\delta e_{t,i}(r_{p,i-1}, \delta e_{t,i-1}), \eta; v_{p}\right] \cdot
\vec{X} \left[v_{p,i}\left(v_{p,i-1}, \delta e_{t,i-1}\right)\right]\right\}\,,
\end{align}
where the functionals $v_{p,i}(v_{p,i-1}, \delta e_{t,i-1})$ and $\delta e_{t,i}(v_{p,i-1}, \delta e_{t,i-1})$ are given in Eqs.~\eqref{vp-map-PN} and~\eqref{e-map-PN}, respectively. The components of the amplitude vectors $\vec{R}^{(-1)}$ are defined recursively in~\ref{recursion}. The time mapping can trivially be constructed from Eq.~\eqref{Torb-PN} via
\begin{align}
\label{t-map-PN}
t_{i} &= t_{i-1} + \frac{2 \pi M}{\left[v_{p,i} \left(v_{p,i-1}, \delta e_{t,i-1}\right)\right]^{3}}  \frac{\left[2 - \delta e_{t,i}\left(v_{p,i-1}, \delta e_{t,i-1}\right)\right]^{3/2}}{\left[\delta e_{t,i} \left(v_{p,i-1}, \delta e_{t,i-1}\right)\right]^{3/2}} 
\nn\\
&\times \left\{1 + \vec{P}\left[\delta e_{t,i}\left(v_{p,i-1}, \delta e_{t,i-1}\right), \eta; v_{p}\right] \cdot \vec{X}\left[v_{p,i} \left(v_{p,i-1}, \delta e_{t,i-1}\right)\right] \right\}\,,
\end{align}
which completes our calculation of the centroid mapping.

We have here {\emph{not}} inserted Eqs.~\eqref{vp-map-PN} and~\eqref{e-map-PN} into Eqs.~\eqref{t-map-PN} and~\eqref{f-map-PN}, and re-expanded about the pericenter velocity being small to determine the time and frequency of the bursts. The reason for this is that such an expansion results in a significant loss of accuracy compared to numerical evolutions. This results from the behavior of the orbital period in the burst model which behaves as 
\begin{equation}
\frac{1}{\delta e_{t,i}^{3/2}} \sim \frac{1}{\left(\delta e_{t,i-1} + A \; v_{p,i-1}^{5}\right)^{3/2}}\,,
\end{equation}
where $A$ is a constant. Since both $\delta e_{t,i-1}$ and $v_{p,i-1}$ are assumed to be simultaneously but independently small, expanding such a function about only one of them would impose an assumption on their ratio that is not justified. 

Finally, we consider the volume mapping of the bursts. Once again, we treat the bursts as boxes in time and frequency with widths defined by Eqs.~\eqref{delta-t} and~\eqref{delta-f}. Hence we simply have to evaluate these expressions within our PN formalism at $(v_{p,i}, e_{t,i})$, thus obtaining
\begin{align}
\delta t_{i} &= \frac{\xi_{t} M \left[2 - \delta e_{t,i}\left(v_{p,i-1}, \delta e_{t,i-1}\right)\right]}{\left[v_{p,i}\left(v_{p,i-1}, \delta e_{t,i-1}\right)\right]^{3}} \left\{1 + \vec{R}\left[\delta e_{t,i}(v_{p,i-1}, \delta e_{t,i-1}), \eta; v_{p}\right] \cdot \vec{X}\left[v_{p,i}\left(v_{p,i-1}, \delta e_{t,i-1}\right)\right]\right\}\,,
\end{align}
\begin{align}
\delta f_{i} &= \frac{\xi_{f} \left[v_{p,i} \left(v_{p,i-1}, \delta e_{t,i-1}\right)\right]^{3}}{2 \pi M \left[2 - \delta e_{t,i} \left(v_{p,i-1}, \delta e_{t,i-1}\right)\right]}  \left\{1 + \vec{R}^{(-1)}\left[\delta e_{t,i}(r_{p,i-1}, \delta e_{t,i-1}), \eta; v_{p}\right] \cdot \vec{X}\left[v_{p,i}\left(v_{p,i-1}, \delta e_{t,i-1}\right)\right]\right\}\,.
\end{align}
Not surprisingly, we see that one set of corrections, specifically the PN corrections to the pericenter distance, are the PN corrections to the frequency and box size mappings. This is exactly like in the ppE burst formalism, where one ppE deformation (with two parameters: $\beta_{\rm ppE}, \bar{b}_{\rm ppE}$) characterized these mappings.

\subsection{Example PN Burst Models}
\label{examples}

We have applied the fully general formalism of Section~\ref{simp} to PN theory, developing a burst model at generic PN order. This arbitrary order model is characterized by four amplitude vector fields $\lambda^{a}_{\rm PN burst} = \left(\vec{P}, \vec{R}, \vec{\cal{V}}, \vec{\cal{D}}\right)$, which can easily be constructed from the PN corrections to the orbital period, pericenter velocity, and rate of change of pericenter velocity and orbital eccentricity. These amplitude vectors are dependent on the set of GR parameters that characterize the system, namely $\lambda^{a}_{\GR} = \left(\delta e_{t}, \eta, ...\right)$. We will now apply this formalism to generate a few example burst models at specific PN orders. There are multiple coordinate systems used to calculate PN quantities. Two that are typically used within the literature are the ADM and modified harmonic coordinates. We will choose to work within the ADM coordinates. The expressions in modified harmonic coordinates can easily be obtained through the appropriate coordinate transformations, which are given for example in Eq.~(7.11) in~\cite{Arun:2007sg}.

\subsubsection{Burst Model at 1PN Order}
We begin by calculating the burst model to 1PN order. Recall that the model has three ingredients: the orbit evolution, the centroid mapping, and the volume mapping. We begin with the orbit evolution, which in our generic order model is given by Eqs.~\eqref{vp-map-PN} and~\eqref{e-map-PN}. There are no 0.5PN order corrections to any of the quantities considered here, so the state vector has only one component, specifically
\begin{equation}
\vec{X} = (v_{p}^{2})\,.
\end{equation}
To achieve a burst model at 1PN order, we simply have to compute the 1PN functions $({\cal{V}}_{2}, {\cal{D}}_{2})$. The functions $({\cal{V}}_{k}, {\cal{D}}_{k})$ are given in general by Eq.~\eqref{calV} and~\eqref{calD}, respectively. Setting $k=2$, these functions become
\begin{align}
{\cal{V}}_{2}(e_{t}, \eta; v_{p}) &= V_{2}(e_{t}, \eta; v_{p}) + P_{2}(e_{t}, \eta; v_{p})\,,
\\
{\cal{D}}_{2}(\delta e_{t}, \eta; v_{p}) &= E_{2}(e_{t}, \eta; v_{p}) + P_{2}(e_{t}, \eta; v_{p})\,.
\end{align}
where $(V_{2}, E_{2}, P_{2})$ are given in~\ref{fields}. Working to ${\cal{O}}(\delta e_{t})$, the orbit evolution becomes
\begin{align}
\label{vp-map-1PN}
\frac{\left(v_{p,i} - v_{p,i-1}\right)_{\rm 1PN}}{\left(v_{p,i} - v_{p,i-1}\right)_{\rm N}} &= 1 + {\cal{V}}_{2}(\delta e_{t,i-1}, \eta; v_{p}) v_{p,i-1}^{2} + {\cal{O}}\left(v_{p,i-1}^{3}\right)
\\
\label{et-map-1PN}
\frac{\left(\delta e_{t,i} - \delta e_{t,i-1}\right)_{\rm 1PN}}{\left(\delta e_{t,i} - \delta e_{t,i-1}\right)_{\rm N}} &= 1 + {\cal{D}}_{2}(\delta e_{t,i-1}, \eta; v_{p}) v_{p,i-1}^{2} + {\cal{O}}\left(v_{p,i-1}^{3}\right)
\end{align}
with
\begin{align}
\label{vp-map-N}
\left(v_{p,i} - v_{p,i-1}\right)_{\rm N} &= - \frac{13 \pi}{96} \eta v_{p, i-1}^{6} \left[1 + \frac{44}{65} \delta e_{t,i-1} + {\cal{O}}\left(\delta e_{t,i-1}^{2}\right) \right]
\\
\label{et-map-N}
\left(\delta e_{t,i} - \delta e_{t,i-1}\right)_{\rm N} &= \frac{85 \pi}{48} \eta v_{p,i-1}^{5} \left[1 + \frac{791}{850} \delta e_{t,i-1}+ {\cal{O}}\left(\delta e_{t,i-1}^{2}\right) \right]
\\
\label{calV-2}
{\cal{V}}_{2}(\delta e_{t,i-1}, \eta; v_{p}) &= -\frac{251}{104} \eta + \frac{8321}{2080} + \delta e_{t,i-1}  \left(\frac{14541}{6760} \eta - \frac{98519}{135200}\right)+ {\cal{O}}(\delta e_{t,i-1}^{2})\,,
\\
\label{calD-2}
{\cal{D}}_{2}(\delta e_{t,i-1}, \eta; v_{p}) &= -\frac{4017}{680} \eta + \frac{4773}{800} + \delta e_{t,i-1} \left(\frac{225393}{144500} \eta - \frac{602109}{340000}\right) + {\cal{O}}(\delta e_{t,i-1}^{2})\,.
\end{align}

Let us now consider the centroid mapping. The evolution of the time centroid of the bursts is trivially given by the orbital period, so to 1PN order
\begin{align}
\frac{\left(t_{i} - t_{i-1}\right)_{\rm 1PN}}{\left(t_{i} - t_{i-1}\right)_{\rm N}} &= 1 + P_{2}\left[\delta e_{t,i}(v_{p,i-1}, \delta e_{t,i-1}), \eta; v_{p}\right] \left[v_{p,i}(v_{p,i-1}, \delta e_{t,i-1})\right]^{2} + {\cal{O}}\left(v_{p,i}^{4}\right)\,,
\end{align}
with
\begin{align}
\left(t_{i} - t_{i-1}\right)_{\rm N} &= P^{\rm N}\left(v_{p,i}, \delta e_{t,i}\right)
\nn \\
&= \frac{2 \pi M}{\left[v_{p,i}(v_{p,i-1}, \delta e_{t,i-1})\right]^{3}} \frac{\left[2 - \delta e_{t,i}(v_{p,i-1}, \delta e_{t,i-1})\right]^{3/2}}{\left[\delta e_{t,i}(v_{p,i-1}, \delta e_{t,i-1})\right]^{3/2}}\,,
\\
P_{2}(\delta e_{t,i}, \eta; v_{p}) &= \frac{3}{2} \eta - \frac{3}{4} + \delta e_{t,i} \left(-\frac{5}{8} \eta + \frac{3}{4}\right) + {\cal{O}}(\delta e_{t,i}^{2})\,,
\end{align}
where $v_{p,i}(v_{p,i-1}, \delta e_{t,i-1})$ and $\delta e_{t,i}(v_{p,i-1}, \delta e_{t,i-1})$ are given by Eqs.~\eqref{vp-map-1PN} and~\eqref{et-map-1PN}, respectively. 

We now move onto the frequency centroid mapping, which is characterized by the functions $R_{k}^{(-1)}$. Using the recursion method is~\ref{recursion}, $R_{2}^{(-1)} = - R_{2}$, and the frequency centroid mapping becomes
\begin{align}
\frac{f_{i}^{\rm PN}}{f_{i}^{\rm N}} &= 1 - R_{2}\left[\delta e_{t,i}(v_{p,i-1}, \delta e_{t,i-1}), \eta; v_{p}\right] \left[v_{p,i}(v_{p,i-1}, \delta e_{t,i-1})\right]^{2} + {\cal{O}}(v_{p,i}^{4})
\end{align}
with
\begin{align}
f_{i}^{\rm N} &= \frac{\left[v_{p,i} \left(v_{p,i-1}, \delta e_{t,i-1}\right)\right]^{3}}{2 \pi M \left[2 - \delta e_{t,i} \left(v_{p,i-1}, \delta e_{t,i-1}\right)\right]}
\\
R_{2}(\delta e_{t,i}, \eta; v_{p}) &= \frac{7}{4} \eta - \frac{5}{2} - \frac{5}{8} \eta \delta e_{t,i} + {\cal{O}}(\delta e_{t,i}^{2})\,.
\end{align}

Finally, we focus on the volume mapping, which is trivially given by the same corrections as the frequency centroid mapping:
\begin{align}
\frac{\delta t_{i}^{\rm 1PN}}{\delta t_{i}^{\rm N}} &= 1 + R_{2}\left[\delta e_{t,i}(v_{p,i-1}, \delta e_{t,i-1}), \eta; v_{p}\right] \left[v_{p,i}(v_{p,i-1}, \delta e_{t,i-1})\right]^{2} + {\cal{O}}(v_{p,i}^{4})\,,
\\
\frac{\delta f_{i}^{\rm 1PN}}{\delta f_{i}^{\rm N}} &= 1 - R_{2}\left[\delta e_{t,i}(v_{p,i-1}, \delta e_{t,i-1}), \eta; v_{p}\right] \left[v_{p,i}(v_{p,i-1}, \delta e_{t,i-1})\right]^{2} + {\cal{O}}(v_{p,i}^{4})
\end{align}
where we have defined
\begin{align}
\delta t_{i}^{\rm N} &= \frac{\xi_{t} M \left[2 - \delta e_{t,i}\left(v_{p,i-1}, \delta e_{t,i-1}\right)\right]}{\left[v_{p,i}\left(v_{p,i-1}, \delta e_{t,i-1}\right)\right]^{3}}
\\
\delta f_{i}^{\rm N} &= \frac{\xi_{f} \left[v_{p,i} \left(v_{p,i-1}, \delta e_{t,i-1}\right)\right]^{3}}{2 \pi M \left[2 - \delta e_{t,i} \left(v_{p,i-1}, \delta e_{t,i-1}\right)\right]}
\end{align}
This completes the burst model at 1PN order.
\subsubsection{Burst Model at 2PN Order}

Let us now calculate the burst model to 2PN order. The state vector has three components corresponding to 1PN, 1.5PN, and 2PN orders, specifically
\begin{equation}
\vec{X} = (v_{p}^{2}, v_{p}^{3}, v_{p}^{4})\,.
\end{equation}
We begin by computing the orbital evolution in the burst model. To 2PN order, the pericenter velocity and eccentricity mappings become
\begin{align}
\label{vp-map-2PN}
\frac{(v_{p,i} - v_{p,i-1})_{\rm 2PN}}{(v_{p,i} - v_{p,i-1})_{\rm N}} &= \frac{(v_{p,i} - v_{p,i-1})_{\rm 1PN}}{(v_{p,i} - v_{p,i-1})_{\rm N}} + {\cal{V}}_{3}(\delta e_{t,i-1}, \eta; v_{p}) v_{p,i-1}^{3} + {\cal{V}}_{4}(\delta e_{t,i-1}, \eta; v_{p}) v_{p,i-1}^{4} 
\nn \\
&+ {\cal{O}}(v_{p,i-1}^{5})
\\
\label{et-map-2PN}
\frac{(\delta e_{t,i} - \delta e_{t,i-1})_{\rm 2PN}}{(\delta e_{t,i} - \delta e_{t,i-1})_{\rm N}} &= \frac{(\delta e_{t,i} - \delta e_{t,i-1})_{\rm 1PN}}{(\delta e_{t,i} - \delta e_{t,i-1})_{\rm N}} + {\cal{D}}_{3}(\delta e_{t,i-1}, \eta; v_{p}) v_{p,i-1}^{3} + {\cal{D}}_{4}(\delta e_{t,i-1}, \eta; v_{p}) v_{p,i-1}^{4} 
\nn \\
&+ {\cal{O}}(v_{p,i-1}^{5})
\end{align}
The Newtonian and 1PN order mappings do not change from the 1PN order model, and they are given in Eqs.~\eqref{vp-map-N}-\eqref{et-map-N} and Eqs.~\eqref{vp-map-1PN}-\eqref{et-map-1PN}, respectively. Generally, the 1.5PN order and 2PN order components of the amplitude fields are given by
\begin{align}
{\cal{V}}_{3}(e_{t}, \eta; v_{p}) &= V_{3}(e_{t}, \eta; v_{p})
\\
{\cal{D}}_{3}(e_{t}, \eta; v_{p}) &= E_{3}(e_{t}, \eta; v_{p})
\\
{\cal{V}}_{4}(e_{t}, \eta; v_{p}) &= V_{4}(e_{t}, \eta; v_{p}) + P_{4}(e_{t}, \eta; v_{p}) + V_{2}(e_{t}, \eta; v_{p}) P_{2}(e_{t}, \eta; v_{p})
\\
{\cal{D}}_{4}(e_{t}, \eta; v_{p}) &= E_{4}(e_{t}, \eta; v_{p}) + P_{4}(e_{t}, \eta; v_{p}) + E_{2}(e_{t}, \eta; v_{p}) P_{2}(e_{t}, \eta; v_{p})
\end{align}
Using the results of~\ref{fields}, we obtain
\begin{align}
{\cal{V}}_{3}(\delta e_{t,i-1}, \eta; v_{p}) &= \frac{3712 \sqrt{3}}{585} + \frac{100864 \sqrt{3}}{12675} \delta e_{t,i-1} + {\cal{O}}(\delta e_{t,i-1}^{2})\,,
\\
{\cal{D}}_{3}(\delta e_{t,i-1}, \eta; v_{p}) &= \frac{10624\sqrt{3}}{3825} + \frac{1098176 \sqrt{3}}{541875} \delta e_{t,i-1} + {\cal{O}}(\delta e_{t,i-1}^{2})\,,
\\
{\cal{V}}_{4}(\delta e_{t,i-1}, \eta; v_{p}) &= \frac{119432023}{6289920} - \frac{1213031}{49920}\eta - \frac{169}{128}\eta^{2} + \delta e_{t,i-1} \left(\frac{29330909}{204422400} + \frac{816679}{202800} \eta 
\right.
\nn \\
&\left.
- \frac{68571}{4160} \eta^{2}\right) + {\cal{O}}(\delta e_{t,i-1}^{3/2})\,,
\\
{\cal{D}}_{4}(\delta e_{t,i-1}, \eta; v_{p}) &= \frac{130397759}{4569600} - \frac{5863719}{108800}\eta + \frac{284687}{10880}\eta^{2} + \delta e_{t,i-1}^{1/2} \left(\frac{45 \sqrt{2}}{32} - \frac{9 \sqrt{2}}{16} \eta\right)
\nn \\
&+ \delta e_{t,i-1} \left(-\frac{26000488883}{2913120000} + \frac{887490277}{46240000}\eta - \frac{16138299}{1156000}\eta^{2}\right) + {\cal{O}}(\delta e_{t,i-1}^{3/2})\,,
\end{align}
where we have used the results of~\cite{Loutrel:2016cdw} to evaluate the tail enhancement factors.

Next, let us consider the time centroid mapping, which at 2PN order is
\begin{align}
\frac{(t_{i} - t_{i-1})_{\rm 2PN}}{(t_{i} - t_{i-1})_{\rm N}} &= \frac{(t_{i} - t_{i-1})_{\rm 1PN}}{(t_{i} - t_{i-1})_{\rm N}} + P_{4}[\delta e_{t,i}(v_{p,i-1}, \delta e_{t,i-1}), \eta; v_{p}] v_{p,i}(v_{p,i-1}, \delta e_{t,i-1})^{4} + {\cal{O}}(v_{p,i}^{5})\,,
\end{align}
where the mappings $v_{p,i}(v_{p,i-1}, \delta e_{t,i-1})$ and $\delta e_{t,i}(v_{p,i-1}, \delta e_{t,i-1})$ are now given by Eqs.~\eqref{vp-map-2PN} and~\eqref{et-map-2PN}, respectively. There is no 1.5PN order correction to these expressions, since $P_{3} = 0$. This is a result of the fact that the orbital period comes from the conservative orbital dynamics, and is thus symmetric under time reversal. Once again, the Newtonian time centroid mapping and 1PN order correction do not change from the 1PN order burst model. The 2PN order correction is characterized solely by $P_{4}(e_{t}, \eta; v_{p})$, which is given in~\ref{fields}. Expanding about $\delta e_{t} \ll 1$, we obtain
\begin{align}
P_{4}(\delta e_{t,i}, \eta; v_{p}) &= - \frac{225}{64} + \frac{237}{64} \eta - \frac{39}{32} \eta^{2} + \delta e_{t,i}^{1/2} \left(\frac{45 \sqrt{2}}{32} - \frac{9 \sqrt{2}}{16} \eta\right) 
\nn\\
&+ \delta e_{t,i} \left(-\frac{135}{64} + \frac{115}{64} \eta + \frac{7}{16} \eta^{2}\right) + {\cal{O}}(\delta e_{t,i}^{3/2})\,.
\end{align}

Finally, consider the frequency centroid and box size mappings, which at 2PN order are
\begin{align}
\frac{f_{i}^{\rm 2PN}}{f_{i}^{\rm N}} &= \frac{f_{i}^{\rm 1PN}}{f_{i}^{\rm N}} + R_{4}^{(-1)}[\delta e_{t,i}(v_{p,i-1}, \delta e_{t,i-1}), \eta; v_{p}]  v_{p,i}(v_{p,i-1}, \delta e_{t,i-1})^{4} + {\cal{O}}(v_{p,i}^{5})\,,
\\
\frac{\delta t_{i}^{\rm 2PN}}{\delta t_{i}^{\rm N}} &=  \frac{\delta t_{i}^{\rm 1PN}}{\delta t_{i}^{\rm N}} + R_{4}[\delta e_{t,i}(v_{p,i-1}, \delta e_{t,i-1}), \eta; v_{p}]  v_{p,i}(v_{p,i-1}, \delta e_{t,i-1})^{4} + {\cal{O}}(v_{p,i}^{5})\,,
\\
\frac{\delta f_{i}^{\rm 2PN}}{\delta f_{i}^{\rm N}} &= \frac{\delta f_{i}^{\rm 1PN}}{\delta f_{i}^{\rm N}} + R_{4}^{(-1)}[\delta e_{t,i}(v_{p,i-1}, \delta e_{t,i-1}), \eta; v_{p}]  v_{p,i}(v_{p,i-1}, \delta e_{t,i-1})^{4} + {\cal{O}}(v_{p,i}^{5})\,,
\end{align}
The Newtonian and 1PN order terms are the same as those in the 1PN order burst model. Using the results of~\ref{recursion}, the field $R_{4}^{(-1)}$ is in general given by
\begin{equation}
R_{4}^{(-1)}(e_{t}, \eta; v_{p}) = -R_{4}(e_{t}, \eta; v_{p}) + R_{2}(e_{t}, \eta; v_{p})^{2}\,.
\end{equation}
Applying the expressions for $(R_{2}, R_{4})$ from~\ref{fields}, we obtain
\begin{align}
R_{4}(\delta e_{t,i}, \eta; v_{p}) &= -\frac{47}{16} + \frac{49}{16} \eta - \frac{17}{16} \eta^{2} + \delta e_{t,i} \left(-\frac{133}{64} + \frac{155}{64} \eta + \frac{13}{32} \eta^{2}\right) + {\cal{O}}(\delta e_{t,i}^{2})\,,
\\
R_{4}^{(-1)}(\delta e_{t,i}, \eta, v_{p}) &= \frac{147}{16} - \frac{189}{16} \eta + \frac{33}{8} \eta^{2} + \delta e_{t,i} \left(\frac{133}{64} + \frac{45}{64} \eta - \frac{83}{32} \eta^{2}\right) + {\cal{O}}(\delta e_{t,i}^{2})\,.
\end{align}
This completes the burst model at 2PN order.
\subsubsection{Burst Model at 3PN Order}
\label{burst-3PN}
Let us now extend the burst model to the current limits of our understanding of eccentric binaries within PN theory, i.e.~to 3PN order. The state vector will now extend to $v_{p}^{6}$, specifically
\begin{equation}
\vec{X} = (v_{p}^{2}, v_{p}^{3}, v_{p}^{4}, v_{p}^{5}, v_{p}^{6})\,.
\end{equation}
At 3PN order, the orbital evolution equations become
\begin{align}
\label{vp-map-3PN}
\frac{(v_{p,i} - v_{p,i-1})_{\rm 3PN}}{(v_{p,i} - v_{p, i-1})_{\rm N}} &= \frac{(v_{p,i} - v_{p,i-1})_{\rm 2PN}}{(v_{p,i} - v_{p, i-1})_{\rm N}}+ {\cal{V}}_{5}(\delta e_{t,i-1}, \eta; v_{p}) v_{p,i-1}^{5} + {\cal{V}}_{6}(\delta e_{t,i-1}, \eta; v_{p}) v_{p,i-1}^{6} 
\nn \\
&+ {\cal{O}}\left(v_{p,i-1}^{7}\right)\,,
\\
\label{e-map-3PN}
\frac{(\delta e_{t,i} - \delta e_{t,i-1})_{\rm 3PN}}{(\delta e_{t,i} - \delta e_{t,i-1})_{\rm N}} &= \frac{(\delta e_{t,i} - \delta e_{t,i-1})_{\rm 2PN}}{(\delta e_{t,i} - \delta e_{t,i-1})_{\rm N}} + {\cal{D}}_{5}(\delta e_{t,i-1}, \eta; v_{p}) v_{p,i-1}^{5} + {\cal{D}}_{6}(\delta e_{t,i-1}, \eta; v_{p}) v_{p,i-1}^{6} 
\nn \\
&+ {\cal{O}}\left(v_{p,i-1}^{7}\right)\,.
\end{align}
The new functions $[{\cal{V}}_{5}, {\cal{V}}_{6}]$ and $[{\cal{D}}_{5}, {\cal{D}}_{6}]$ give the coefficients of the 2.5PN and 3PN order corrections of the orbital evolutions. In terms of the components of the amplitude vector fields $[\vec{V}, \vec{E}, \vec{P}]$, they are given by
\begin{align}
{\cal{V}}_{5}(e_{t}, \eta; v_{p}) &= V_{5}(e_{t}, \eta; v_{p}) + V_{3}(e_{t}, \eta; v_{p}) P_{2}(e_{t}, \eta; v_{p})\,,
\\
{\cal{D}}_{5}(e_{t}, \eta; v_{p}) &= E_{5}(e_{t}, \eta; v_{p}) + E_{3}(e_{t}, \eta; v_{p}) P_{2}(e_{t}, \eta; v_{p})\,,
\\
{\cal{V}}_{6}(e_{t}, \eta; v_{p}) &= V_{6}(e_{t}, \eta; v_{p}) + P_{6}(e_{t}, \eta; v_{p}) + V_{2}(e_{t}, \eta; v_{p}) P_{4}(e_{t}, \eta; v_{p}) 
\nn \\
&+ V_{4}(e_{t}, \eta; v_{p}) P_{2}(e_{t}, \eta; v_{p})\,,
\\
{\cal{D}}_{6}(e_{t}, \eta; v_{p}) &= E_{6}(e_{t}, \eta; v_{p}) + P_{6}(e_{t}, \eta; v_{p}) + E_{2}(e_{t}, \eta; v_{p}) P_{4}(e_{t}, \eta; v_{p}) 
\nn \\
&+ E_{4}(e_{t}, \eta; v_{p}) P_{2}(e_{t}, \eta; v_{p})\,,
\end{align}
where we have used the fact that $P_{3}(e_{t}, \eta; v_{p}) = 0 = P_{5}(e_{t}, \eta; v_{p})$. Using the results of~\ref{fields}, we find for the 2.5PN order functions
\begin{align}
{\cal{V}}_{5}(\delta e_{t,i-1}, \eta; v_{p}) &= - \frac{128272 \sqrt{3}}{4095} - \frac{4832 \sqrt{3}}{117} \eta + \nu_{0} \pi + \frac{1748 \sqrt{6}}{65} \delta e_{t,i-1}^{1/2} 
\nn \\
&+ \delta e_{t,i-1} \left(- \frac{30641528 \sqrt{3}}{266175} - \frac{1183488 \sqrt{3}}{29575} \eta + \nu_{1} \pi \right) + {\cal{O}}(\delta e_{t,i-1}^{3/2})\,,
\\
{\cal{D}}_{5}(\delta e_{t,i-1}, \eta; v_{p}) &= \frac{13072 \sqrt{3}}{8925} - \frac{241664 \sqrt{3}}{8925} \eta + \rho_{0} \pi + \frac{4544 \sqrt{6}}{425} \delta e_{t,i-1}^{1/2} 
\nn \\
&+ \delta e_{t,i-1} \left(- \frac{81300056 \sqrt{3}}{3793125} - \frac{52270208 \sqrt{3}}{3793125} \eta + \rho_{1} \pi \right) + {\cal{O}}(\delta e_{t,i-1}^{3/2})\,,
\end{align}
where we have used the results of~\cite{Loutrel:2016cdw} and neglected the 2.5PN memory terms. The constants $[\nu_{0}, \nu_{1}, \rho_{0}, \rho_{1}]$ depend on the coefficients of the Pad\'{e} approximants created for the 2.5PN order tail enhancement factors $[\psi(e_{t}), \tilde{\psi}(e_{t})]$ in~\cite{Loutrel:2016cdw}. The exact rational form of the coefficients are too lengthy to provide here. We simply give their numeric values, which are
\begin{align}
\nu_{0} &= 34.82829720\,,
\qquad
\nu_{1} = -38.97374189\,,
\\
\rho_{0} &= 11.90237615\,,
\qquad
\rho_{1} = -36.89484102\,,
\end{align}
For the 3PN order functions, we find 
\begin{align}
{\cal{V}}_{6}(\delta e_{t,i-1}, \eta; v_{p}) &= \frac{48102359171}{402554880} + \frac{385 \pi^{2}}{128} + \frac{1177 {\rm ln}(2)}{64} + \frac{1177 {\rm ln}(3)}{256} 
\nn \\
&- \left(\frac{508363\pi^{2}}{266240} + \frac{80844193}{430080}\right) \eta + \frac{3379743}{53248} \eta^{2} + \frac{543189}{13312} \eta^{3} - \frac{1177}{256} {\rm ln}(v_{p,i-1}^{2})
\nn \\
&+ \delta e_{t,i-1} \left[\frac{1690426235921}{26166067200} + \frac{48839 \pi^{2}}{8320} + \frac{746539 {\rm ln}(2)}{20800} + \frac{746539 {\rm ln}(3)}{83200}
\right.
\nn \\
&\left.
- \left(\frac{112925149}{83865600} + \frac{80684263 \pi^{2}}{17305600}\right) \eta - \frac{753359873}{10383360} \eta^{2} + \frac{61283003}{865280} \eta^{3}
\right.
\nn \\
&\left.
-\frac{746539}{83200} {\rm ln}(v_{p,i-1}^{2})\right] + {\cal{O}}\left(\delta e_{t,i-1}^{3/2}\right)\,,
\\
{\cal{D}}_{6}(\delta e_{t,i-1}, \eta; v_{p}) &= \frac{7318191053}{51609600} + \frac{10549 \pi^{2}}{10880} + \frac{161249 {\rm ln}(2)}{27200} + \frac{161249 {\rm ln}(3)}{108800}  
\nn \\
&- \left(\frac{8119255961}{21934080} + \frac{155561 \pi^{2}}{1740800}\right) \eta + \frac{11789862391}{36556800} \eta^{2} - \frac{9152141}{87040} \eta^{3} - \frac{161249}{108800} {\rm ln}(v_{p,i-1}^{2}) 
\nn \\
&+ \delta e_{t,i-1}^{1/2} \left[\frac{64557 \sqrt{2}}{5120} - \left(\frac{4120619 \sqrt{2}}{217600} - \frac{123 \pi^{2} \sqrt{2}}{4096}\right)  \eta + \frac{13437 \sqrt{2}}{2720} \eta^{2} \right] 
\nn \\
&+ \delta e_{t,i-1} \left[- \frac{11487739123}{552960000} + \frac{5805723 \pi^{2}}{4624000} + \frac{88744623 {\rm ln}(2)}{11560000} + \frac{88744623 {\rm ln}(3)}{46240000} 
\right.
\nn \\
&\left.
+ \left(\frac{639985247281}{6991488000} + \frac{13567261 \pi^{2}}{92480000} \right)\eta - \frac{689800811001}{5178880000} \eta^{2} + \frac{1557091039}{18496000} \eta^{3} 
\right.
\nn \\
&\left.
- \frac{88744623}{46240000} {\rm ln}(v_{p,i-1}^{2}) \right] + {\cal{O}}(\delta e_{t,i-1}^{3/2})\,.
\end{align}
This completes the orbital evolution to 3PN order.

The time centroid mapping at 3PN order becomes
\begin{align}
\label{t-map-3PN}
\frac{(t_{i} - t_{i-1})_{\rm 3PN}}{(t_{i} - t_{i-1})_{\rm N}} &= \frac{(t_{i} - t_{i-1})_{\rm 2PN}}{(t_{i} - t_{i-1})_{\rm N}} + P_{6}[\delta e_{t,i}(v_{p,i-1}, \delta e_{t,i-1}), \eta; v_{p}]  v_{p,i}^{6} + {\cal{O}}(v_{p,i}^{7})\,,
\end{align}
where once again there is no 2.5PN order corrections since the orbital period comes from the conservative orbital dynamics. The 3PN order function $P_{6}(e_{t}, \eta; v_{p})$ is given in~\ref{fields}. Expanding about $\delta e_{t} \ll 1$, we obtain
\begin{align}
P_{6}(\delta e_{t,i}, \eta; v_{p}) &= - \frac{2821}{256} + \left(\frac{2123}{128} + \frac{3 \pi^{2}}{16} \right) \eta - \frac{1377}{128} \eta^{2} + \frac{73}{32} \eta^{3} + \delta e_{t,i}^{1/2} \left[ \frac{405 \sqrt{2}}{128} 
\right.
\nn \\
&\left.
- \left(\frac{607 \sqrt{2}}{128} - \frac{123 \pi^{2} \sqrt{2}}{4096}\right) \eta + \frac{99 \sqrt{2}}{128} \eta^{2} \right] + \delta e_{t,i} \left[- \frac{213}{32} + \left(\frac{591}{128} + \frac{885 \pi^{2}}{2048} \right) \eta 
\right.
\nn \\
&\left.- \frac{1117}{512} \eta^{2} - \frac{399}{256} \eta^{3}\right] + {\cal{O}}(\delta e_{t,i}^{3/2})\,.
\end{align}

Finally, the frequency and box widths mappings at 3PN order are
\begin{align}
\label{f-map-3PN}
\frac{f_{i}^{\rm 3PN}}{f_{i}^{\rm N}} &= \frac{f_{i}^{\rm 2PN}}{f_{i}^{\rm N}} + R_{6}^{(-1)}[\delta e_{t,i}(v_{p,i-1}, \delta e_{t,i-1}), \eta; v_{p}]  v_{p,i}(v_{p,i-1}, \delta e_{t,i-1})^{6} + {\cal{O}}(v_{p,i}^{7})\,,
\\
\frac{\delta t_{i}^{\rm 3PN}}{\delta t_{i}^{\rm N}} &= \frac{\delta t_{i}^{\rm 2PN}}{\delta t_{i}^{\rm N}} + R_{6}[\delta e_{t,i}(v_{p,i-1}, \delta e_{t,i-1}), \eta; v_{p}]  v_{p,i}(v_{p,i-1}, \delta e_{t,i-1})^{6} + {\cal{O}}(v_{p,i}^{7})\,,
\\
\frac{\delta f_{i}^{\rm 3PN}}{\delta f_{i}^{\rm N}} &= \frac{\delta f_{i}^{\rm 2PN}}{\delta f_{i}^{\rm N}} + R_{6}^{(-1)}[\delta e_{t,i}(v_{p,i-1}, \delta e_{t,i-1}), \eta; v_{p}]  v_{p,i}(v_{p,i-1}, \delta e_{t,i-1})^{6} + {\cal{O}}(v_{p,i}^{7})\,,
\end{align}
where the functions $R_{6}^{(-1)}(e_{t}, \eta; v_{p})$ is
\begin{align}
R_{6}^{(-1)}(e_{t}, \eta; v_{p}) &= -R_{6}(e_{t}, \eta; v_{p}) + 2 R_{2}(e_{t}, \eta; v_{p}) R_{4}(e_{t}, \eta; v_{p}) - R_{2}(e_{t}, \eta; v_{p})^{3}\,.
\end{align}
Using the results in~\ref{fields}, we obtain
\begin{align}
R_{6}(\delta e_{t,i}, \eta; v_{p}) &= -\frac{305}{32} + \left(\frac{3131}{192} + \frac{11 \pi^{2}}{128} \right) \eta - \frac{19}{2} \eta^{2} + \frac{67}{32} \eta^{3} + \delta e_{t,i} \left[-\frac{829}{128} + \left(\frac{2245}{256} + \frac{97 \pi^{2}}{512}\right) \eta 
\right.
\nn \\
&\left.
- \frac{333}{128} \eta^{2} - \frac{47}{32} \eta^{3} \right] + {\cal{O}}(\delta e_{t,i}^{3/2})\,,
\\
R_{6}^{(-1)}(\delta e_{t,i}, \eta; v_{p}) &= \frac{1275}{32} - \left(\frac{14345}{192} + \frac{11 \pi^{2}}{128} \right) \eta + \frac{97}{2} \eta^{2} - \frac{715}{64}\eta^{3} + \delta e_{t,i} \left[\frac{2159}{128} - \left(\frac{3267}{256} + \frac{97 \pi^{2}}{512}\right) \eta 
\right.
\nn \\
&\left.
- \frac{179}{16} \eta^{2} + \frac{1275}{128} \eta^{3}\right] + {\cal{O}}(\delta e_{t,i}^{3/2})\,.
\end{align}
This completes the burst model at 3PN order.

\section{Properties of the PN Burst Model} 

With the burst model complete to 3PN order, we complete this paper with some results that describe properties of the model. We begin by discussing the accuracy of the burst model when compared to numerical evolutions of the PN radiation reaction equations. Finally, we discuss a previously unreported phenomenon associated with the evolution of the pericenter velocity under radiation reaction.

\subsection{Accuracy of the Burst Model}
\label{accuracy}

The burst model is meant to be an accurate representation of GW bursts emitted by highly elliptic binaries in nature. Further, since this model is designed to be used as a prior in data analysis for detecting such systems, it is paramount that we characterize the accuracy of the model. The ideal test of such an analytic model would be to compare the time of arrival and frequency of eccentric bursts from a numerical relativity simulation to the those from the burst model. However, there are currently no accurate numerical relativity waveforms for the highly elliptic systems considered here. Even the second best comparison, the same as above but with accurate PN waveforms, is also currently inapplicable due to the lack of such waveforms. With the two most ideal tests out of reach, we are left with comparing the burst model to the orbital evolution of binary systems (instead of their associated waveforms) under PN radiation reaction. Such a comparison allows us to gauge the accuracy of the approximations used to construct the burst model, as well as estimate the typical error we can expect when comparing to physically accurate waveform models.

We begin by describing the method through which we obtain the numerical evolution. Ideally, the equations we would want to numerically evolve are $\langle \dot{e}_{t} \rangle (v_{p}, e_{t})$ and $\langle \dot{v}_{p} \rangle (v_{p}, e_{t})$. However, as we will explain in Sec.~\ref{brake}, there is always a point $(v_{p}, e_{t})$ where $\langle \dot{v}_{p} \rangle =0$ during the inspiral, which numerical routines will have difficulty integrating past. An alternative approach is to use a parameterization of the equations that does not present this behavior, e.g.~$\langle \dot{e}_{t} \rangle (x, e_{t})$ and $\langle \dot{x} \rangle (x,e_{t})$. The expression for $\langle \dot{e}_{t} \rangle (x, e_{t})$ to 3PN order, neglecting memory contributions, is provided in Eqs.~(6.18)-(6.19),~(6.22), and~(6.25) in~\cite{Arun:2009mc}. To obtain the expression for $\langle \dot{x} \rangle (x, e_{t})$ to 3PN order, we follow the method detailed in~\ref{fields} for $\langle \dot{v}_{p} \rangle$, which we summarize here. We begin by obtaining an expression for $x(\epsilon, j)$ by inverting Eq.~(6.5) in~\cite{Arun:2009mc}. We then take a time derivative and apply the chain rule, using the 3PN order expressions for the energy flux~\cite{Arun:2007sg} and the angular momentum flux~\cite{Arun:2009mc}. We expand the resulting expression in $x$ to obtain $\langle \dot{x} \rangle (x, e_{t})$. 

For our numerical evolutions, we integrate the equations $\langle \dot{x} \rangle (x, e_{t})$ and $\langle \dot{e}_{t} \rangle (x, e_{t})$ including all of the instantaneous and tail contributions to 3PN order. For the tail enhancement factors, we use the analytic expressions provided in~\cite{Loutrel:2016cdw}. The initial conditions for the evolutions are set to guarantee the initial eccentricity is $e_{t,0} = 0.9$ and the initial GW frequency is $f_{\rm GW,0} = 10 {\rm Hz}$, i.e.~we use these initial conditions to solve for the initial value of $v_{p}$ using $f_{\rm GW}(v_{p},e_{t})$, which is provided in~\ref{fields}. We then use the expression $v_{p}(x, et)$, which is obtained from the 3PN extension of Eq.~\eqref{eq:vp} with Eq.~(7.10) in~\cite{Arun:2007sg}, to obtain the initial value of $x$. For the three systems we study, a $(1.4,1.4) M_{\odot}$ NSNS binary, a $(1.4,10) M_{\odot}$ NSBH binary, and a $(10,10) M_{\odot}$ BHBH binary, the initial conditions are listed in Table~\ref{ic}. With the initial conditions set, we  numerically integrate the equations using the $\textit{NDSolve}$ routine in $\texttt{Mathematica}$ until we reach the time when
\begin{equation}
x_{f} = \frac{1}{2} \left(\frac{1 - e_{t}^{2}}{3 + e_{t}}\right)\,,
\end{equation}
which denotes the maximum value of $x$ for which test particle orbits are stable around a Schwarzschild black hole, i.e.~we require that $p > 2M (3 + e_{t})$, where $p$ is the semi-latus rectum of the orbit and we have used the Newtonian relation $x = (M/p) (1 - e_{t}^2)$. Beyond this point, we consider the inspiral to be formally over and to use the burst model one would have to extend it to include merger and ringdown. 

{\renewcommand{\arraystretch}{1.2}
\begin{table}
\centering
\begin{centering}
\begin{tabular}{cccccc}
\hline
\hline
\noalign{\smallskip}
	 $ \text{System} $  & $ m_{1}[M_{\odot}] $ & $ m_{2}[M_{\odot}] $ & $ e_{t,0} $ & $ x_{0} $ & $ 1/x_{0} $ \\ 
\hline
\noalign{\smallskip}
	$ \text{NSNS} $ & $ 1.4 $ & $ 1.4 $  & $ 0.9 $ & $ 7.35 \times 10^{-4} $ & $ 1360 $ \\
	$ \text{NSBH} $ & $ 1.4 $ & $ 10 $ & $ 0.9 $ & $ 1.85 \times 10^{-3} $ & $ 541 $ \\
	$ \text{BHBH} $ & $ 10 $ & $ 10 $ & $ 0.9 $ & $ 2.67 \times 10^{-3} $ & $ 375 $ \\
\noalign{\smallskip}	
\hline
\hline
\end{tabular}
\end{centering}
\caption{\label{ic} Initial values of the PN expansion parameter $x$ for the set of compact binary systems studied. The values are obtained by requiring the initial GW frequency to be $10 {\rm Hz}$. The final column provides an estimate of the semi-major axis of the binary, since $a_{r} = M/x + {\cal{O}}(1)$ in PN theory.}
\end{table}}

For the comparison to the burst model, we use $x(t)$ and $e_{t}(t)$ to construct the pericenter velocity as a function of time $v_{p}(t)$, which we then use with the results of~\ref{fields} to obtain the orbital period and GW frequency as a function of time, specifically $P(t)$ and $f_{\rm GW}(t)$. To compute the values of these in the burst model to 3PN order, we start the model with the same initial conditions used for the numerical evolution. Once $(v_{p,0}, e_{t,0})$ are specified, all future $(v_{p,i}, e_{t,i})$ are determined from Eqs.~\eqref{vp-map-3PN}-\eqref{e-map-3PN}. From here, the orbital period and the GW frequency are determined in the burst model from Eqs.~\eqref{t-map-3PN} and~\eqref{f-map-3PN}.

\begin{figure*}[ht]
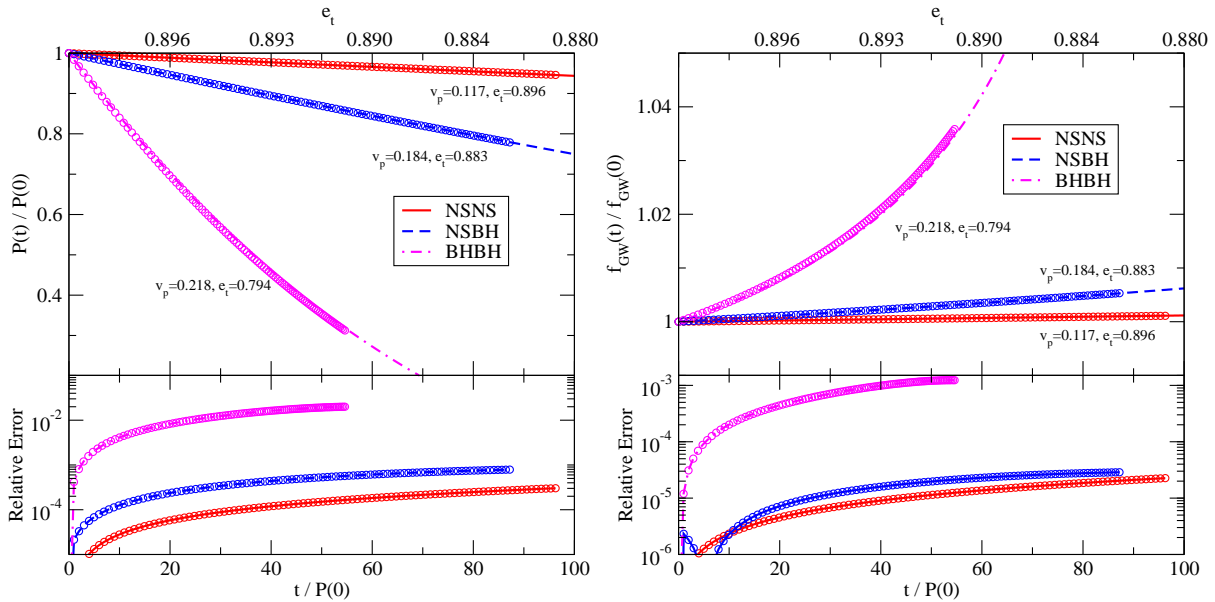

\includegraphics[clip=true,scale=0.34]{TorbComparePN.eps}
\includegraphics[clip=true,scale=0.34]{fGWComparePN.eps}
\caption{\label{compare} Top panel: Comparison of the orbital period $P$ and GW frequency $f_{\rm GW}$ relative to their initial values as functions of time (in units of the initial orbital period) in the burst model (circles) and the numerical evolution (lines). The values of the pericenter velocity and time eccentricity next to each line provide the values during the 100th burst. The labels on the top axis give the value of the time eccentricity for the corresponding time for the NSBH binary. Bottom panel: Relative error between the burst model and the numerical evolutions for the orbital period and GW frequency.}
\end{figure*}

Figure~\ref{compare} shows the orbital period and GW frequency as functions of time in the burst model and the numerical evolution, as well as the relative error between the two. The relative error increases as time increases, but typically the error remains below $1\%$ for the first one hundred bursts. The reason the error increases is twofold. First, the eccentricity decreases as the binary inspirals due to the loss of energy and angular momentum by GW emission. The burst model uses an expansion about $\delta e_{t} \ll 1$, and it is thus most accurate in this regime. This error can be improved by going to higher order in $\delta e_{t}$ within the burst model if one wishes. 

The second reason for the increasing error is that as the binary inspirals, the GW power becomes smeared over more of the orbit. As a result, the binary's evolution resembles less a set of discrete steps. The burst model, which is only valid when $\delta e_{t} \ll 1$, hinges on the osculating behavior of highly eccentric orbits. This error is more difficult to control, but one way of improving it would be to match the evolution in the burst model to an evolution when the eccentricity is small. However, in this paper, we are only interested in highly elliptic orbits where this matching is unnecessary. Regardless, as the figure shows, the error between the burst model and the numerical evolution is sufficiently small that we can begin to test the burst model in idealized data analysis scenarios. 

%
\subsection{Pericenter Braking}
\label{brake}

Let us begin by recalling that within our generic PN burst model, the change of pericenter velocity to Newtonian order is given by
\begin{align}
\label{vp-dot}
\langle \dot{v}_{p} \rangle &= \frac{32}{5} \frac{\eta}{M} v_{p}^{9} \frac{\left(1 - e_{t}\right)^{3/2}}{\left(1 + e_{t}\right)^{15/2}} V_{\rm N}(e_{t}) + {\cal{O}}(v_{p}^{11})
\end{align}
where the function $V_{\rm N}(e_{t})$ is 
\begin{equation}
V_{\rm N}(e_{t}) = 1 - \frac{13}{6} e_{t} + \frac{7}{8} e_{t}^{2} - \frac{37}{96} e_{t}^{3}\,.
\end{equation} 
Notice from Eq.~\eqref{VN}, which provides $V_{\rm N}(\delta e_{t})$, that to first order in $\delta e_{t}$, the above expression is negative and $v_{p}$ is thus decreasing. This seems counterintuitive considering what we know about quasi-circular binaries, i.e.~as the orbital separation $r$ decreases, the orbital velocity $v$ increases, since $v$ and $r$ are inversely related by Kepler's third law. 

This behavior becomes more confusing when we consider the apocenter velocity $v_{a}$. Just as we can calculate $\langle \dot{v}_{p} \rangle$ using the method detailed in~\ref{fields}, we may also compute $\langle \dot{v}_{a} \rangle$. Following this method, and working to Newtonian order, we have
\begin{equation}
\label{va-dot}
\langle \dot{v}_{a} \rangle = \frac{32}{5} \frac{\eta}{M} v_{p}^{9} \frac{\left(1 - e_{t}\right)^{3/2}}{\left(1 + e_{t}\right)^{15/2}} V_{\rm N}(-e_{t}) + {\cal{O}}\left(v_{p}^{11}\right)\,.
\end{equation}
Notice that this expression depends on $V_{\rm N}(-e_{t})$, which is always positive. The apocenter velocity is thus always increasing as the binary inspirals, just as we would expect from quasi-circular binaries. As a result, the pericenter and apocenter velocities have very different behavior depending on the eccentricity of the system.

Let us try to understand this counter-intuitive behavior. The function $V_{\rm N}(e_{t})$ is a third order polynomial in eccentricity with an oscillating sign and with the coefficient of the ${\cal{O}}(e_{t})$ term greater than unity. This means that there will be a critical point $e_{t,{\rm crit}}$ where the function is zero, $\langle \dot{v}_{p} \rangle(e_{t,{\rm crit}}) = 0$, and due to the aforementioned behavior of the coefficients $e_{t,{\rm crit}} < 1$. Let us solve for this critical point. To Newtonian order we find
\begin{equation}
e_{t,{\rm crit}} = e_{t,{\rm crit}}^{\rm N} \equiv \frac{28}{37} - \frac{2}{111} \sigma + \frac{2672}{37} \sigma^{-1}\,, 
\end{equation}
where we have defined
\begin{equation}
\sigma =  \left(67770 + 222 \sqrt{1399593}\right)^{1/3}\,.
\end{equation}
The Newtonian expression for the critical eccentricity evaluates to $e_{\rm crit}^{\rm N} \approx 0.5557306$. Such a critical point also exists at 1PN order, except that now it is a function of the mass ratio and the pericenter velocity:
\begin{equation}
\label{etcrit-1PN}
e_{t,{\rm crit}} = e_{t,{\rm crit}}^{\rm N} + e_{t,{\rm crit}}^{\rm 1PN}(\eta) \; v_{p}^{2}\,,
\end{equation}
where we have defined
\begin{align}
\label{etcrit-1PN-exact}
e_{t,{\rm crit}}^{\rm 1PN}(\eta) &= \frac{1}{\sigma \left(2 \sigma^2-195 \sigma-8016\right)^2 \left(\sigma^4-4008 \sigma^2 + 16064064\right)} \left[\frac{10112970188463538176}{37} \eta 
\right.
\nn \\
&\left.
-\frac{2775205875922795266048}{9583} + \left(\frac{149499372172271616}{37} \eta - \frac{31076515350417788928}{9583}\right) \sigma 
\right.
\nn \\
&\left.
+ \left(-\frac{1150598736488448}{37} \eta-\frac{11994925296964608}{9583}\right) \sigma^2 
\right.
\nn \\
&\left.
+ \left(\frac{62455763578752}{37} \eta - \frac{30800598698771712}{9583}\right) \sigma^3 
\right.
\nn \\
&\left.
+ \left(-\frac{383487115344}{37} \eta + \frac{351826743539100}{9583}\right) \sigma^4
\right.
\nn \\
& \left.
+ \left(-\frac{46179281106}{37} \eta + \frac{35571374119917}{19166}\right) \sigma^5 + \left(\frac{95680418}{37} \eta - \frac{175562247275}{19166}\right) \sigma^6 
\right.
\nn \\
&\left.
+ \left(\frac{3887918}{37} \eta - \frac{1917360308}{9583}\right) \sigma^7 + \left(\frac{53612}{111} \eta + \frac{558902}{28749}\right) \sigma^8 + \left(\frac{1738}{111} \eta - \frac{361279}{28749}\right) \sigma^9 
\right.
\nn \\
&\left.
+ \left(-\frac{88}{333} \eta + \frac{24149}{86247}\right) \sigma^{10}\right]
\end{align}
This function evaluates to $e_{t,\rm crit}^{\rm 1PN} \approx 0.5557306 - (0.06536872 \eta + 0.3457145) v_{p}^{2}$. The overall effect of the 1PN term is to decrease the value of the Newtonian critical point, but there is no value of $v_{p} < 1$ or $\eta \in (0,1/4)$ for which $e_{t,\rm crit} = 0$ at 1PN order.

\begin{figure}[ht]
\centering
\includegraphics[clip=true,scale=0.65]{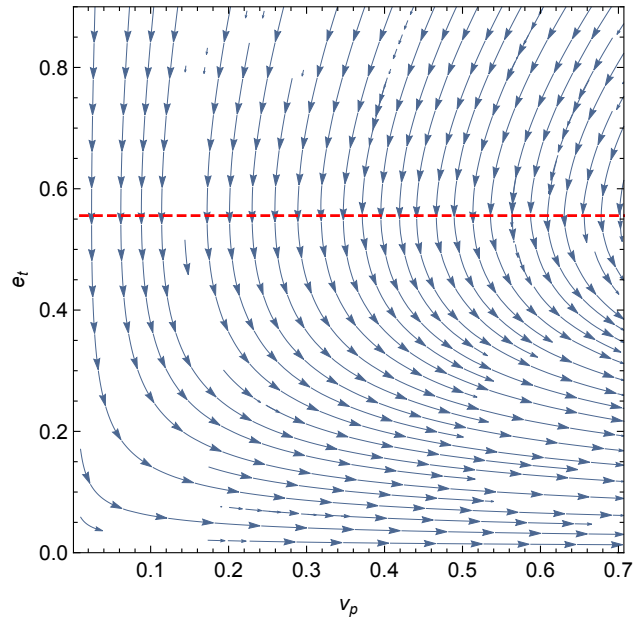}
\caption{\label{braking-N} Plot of the streamlines of ($10^{4} \langle \dot{v}_{p} \rangle, 10^{3} \langle \dot{e}_{t} \rangle$) at Newtonian order. The arrows on the streamlines only indicate the direction of the flow, not the magnitude. The red dashed line displays the value of the critical eccentricity where $\langle \dot{v}_{p} \rangle = 0$ at Newtonian order. Above the critical eccentricity, the streamlines point to the left as shown in the burst model, while below, they point to the right, as is expected for quasi circular binaries.}
\end{figure}
\begin{figure*}[ht]
\includegraphics[clip=true,scale=0.65]{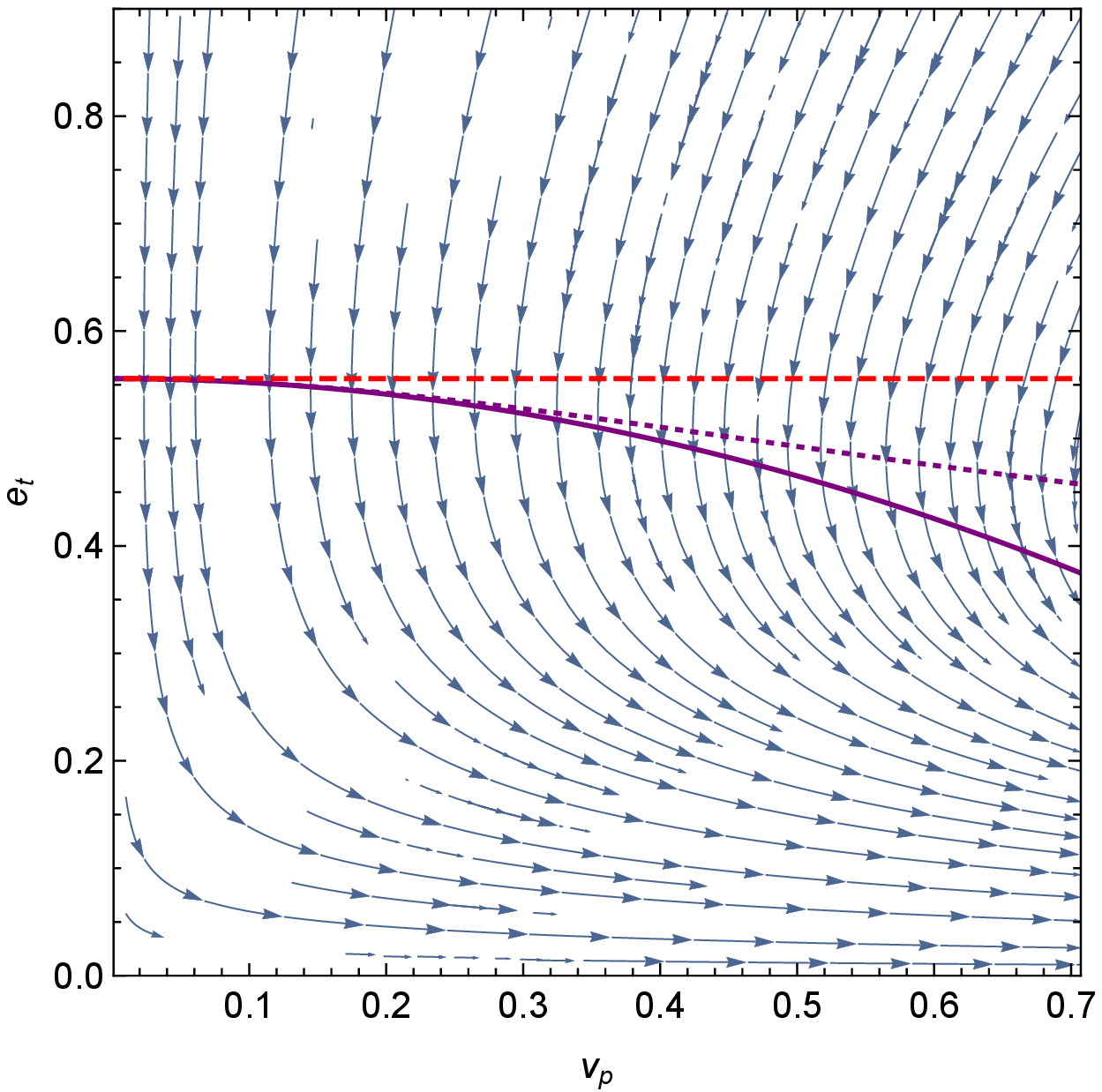}
\includegraphics[clip=true,scale=0.66]{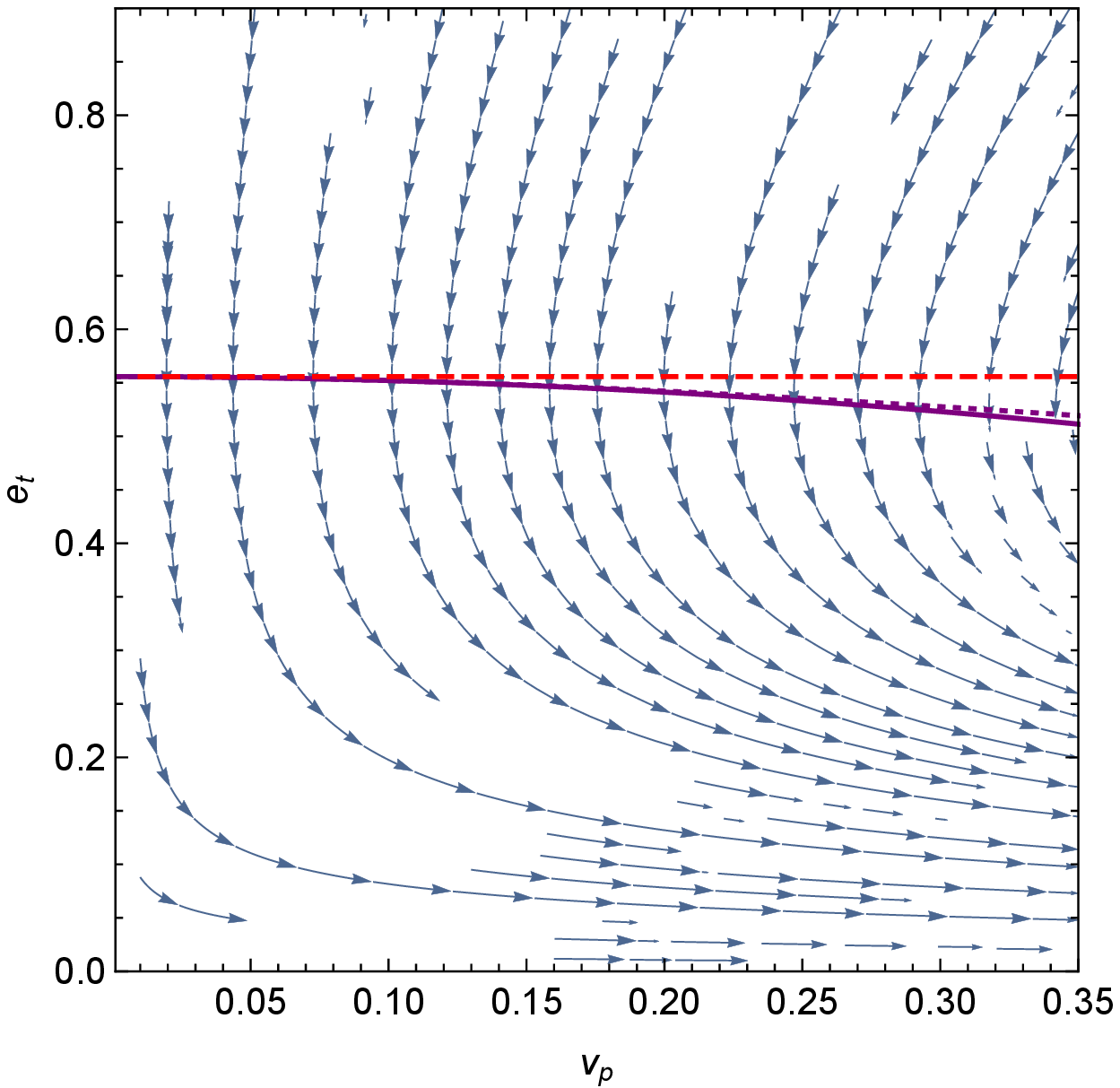}
\caption{\label{braking-PN} Left: Plot of the streamlines of ($10^{4} \langle \dot{v}_{p} \rangle, 10^{3} \langle \dot{e}_{t} \rangle$) at 1PN order. The dotted line displays the value of the critical eccentricity where $\langle \dot{v}_{p} \rangle = 0$ at 1PN order, as determined numerically, while the dashed line is the same result at Newtonian order. The solid line displays the analytic result of the critical eccentricity given in Eqs.~\eqref{etcrit-1PN}-\eqref{etcrit-1PN-exact}. Right: A zoom in of the plot on the left for the region $v_{p} = (0, 0.35)$.}
\end{figure*}

Why does this behavior occur physically? The answer to this question lies in circularization. As the binary inspirals, energy and angular momentum are radiated away in such a way that the orbital eccentricity decreases, making the binary more and more circular. For quasicircular binaries, $v_{a} = v_{p} + {\cal{O}}(e_{t})$, but for highly-elliptic binaries, $v_{p} \gg v_{a}$. As a highly-elliptical binary inspirals, $v_{p}$ and $v_{a}$ will approach the same value since the eccentricity approaches zero. However, if the eccentricity is above the critical value, the two velocities will not approach the same value if they are both increasing initially. Instead, circularization causes pericenter to \emph{brake} when the eccentricity is above $e_{t,{\rm crit}}$, so that $v_{p}$ can approach $v_{a}$. In fact, one can easily show from Eqs.~\eqref{vp-dot} and~\eqref{va-dot}, that $v_{a}$ and $v_{p}$ obey the following conservation law at Newtonian order:
\begin{equation}
V_{\rm N}(-e_{t}) \langle \dot{v}_{p} \rangle - V_{\rm N}(e_{t}) \langle \dot{v}_{a} \rangle = 0\,.
\end{equation}

To further display this behavior, we plot the streamlines of $(\langle \dot{v}_{p} \rangle, \langle \dot{e}_{t} \rangle)$\footnote{We have rescaled the values of $\langle \dot{v}_{p} \rangle$ and $\langle \dot{e}_{t} \rangle$ in these plots to exemplify the behavior of the streamlines. This does not changes the results of this section.} at Newtonian order in Fig~\ref{braking-N} and at 1PN order in Fig.~\ref{braking-PN}. Notice that in both plots, systems with values of $(v_{p}, e_{t})$ above the critical value of the eccentricity, which is represented by the dashed line in Fig.~\ref{braking-N} and the dotted line in Fig.~\ref{braking-PN}, display the pericenter braking behavior that appears in the burst model. On the other hand, the pericenter velocity for systems below the critical eccentricity is always increasing. 

This pericenter braking behavior is not a property of the burst model \emph{per se}, but rather it is inherited from the PN radiation-reaction equations. One may worry that this pericenter braking behavior may disappear if treating the problem exactly (for example, through a numerical treatment). The right panel of Fig.~\ref{braking-PN}, however, shows a zoom of the streamlines at small velocities, where we see that the braking behavior persists. We thus conclude that it is unlikely that pericenter braking is a artifact of the PN expansion. 

Figure~\eqref{braking-PN} also allows us to compare the critical eccentricity computed at Newtonian order, at 1PN order and numerically. The latter is obtained by solving the 1PN expression for $\langle \dot{v}_{p}(e_{t}) \rangle = 0$ to find $e_{t,\rm crit}$. As expected, the numerical inversion disagrees with the its 1PN expansion at high velocities. We notice, however, that the 1PN expression is closer to the numerical inversion than the Newtonian expression is. If the numerical inversion is correct, then this implies the 1PN expansion of $e_{t,\rm crit}$ given in Eqs.~\eqref{etcrit-1PN}-\eqref{etcrit-1PN-exact} has a larger regime of validity than its Newtonian counterpart.  


Finally, it is important to note that while the pericenter velocity has this unique behavior, the GW frequency and the PN parameter $x$ are both monotonically increasing, and the time eccentricity is monotonically decreasing, throughout the inspiral of the binary. In the circular case, there is a one-to-one mapping between the orbital velocity and the GW frequency, and since the orbital velocity is a monotonic function, so is the frequency. For generic eccentric inspirals, the frequency depends on both the pericenter velocity (or alternatively $x$) and the time eccentricity in such a way that it is also monotonic.

\section{Discussion}
\label{discussion}

We have constructed a generic PN order burst model. This model is characterized by four amplitude vector fields $(\vec{P}, \vec{R}, \vec{\cal{V}}, \vec{\cal{D}})$, which depend on the orbital period, pericenter distance, and rates of change of pericenter velocity and orbital eccentricity, respectively. While these quantities are not typically reported within the literature, they can be easily calculated from the quantities that are. Thus, the formalism presented here provides a formulaic means of generating burst models to any PN order. We have then applied this formalism to calculate the burst model out to the current limit to which we can compute PN quantities for eccentric binaries, i.e.~3PN order.

One direction of future research is to relax some of the assumptions used to develop this formalism. For example, we have approximated the compact objects as non-spinning point particles, which is appropriate if we are considering non-spinning BHs. However, BHs in the universe are generally considered to be spinning, while on the other hand, NSs are not well approximated by point particles. NSs will typically have small spins, however the inclusion of finite size effects and tidal perturbations would be necessary to effectively model highly elliptic NS binaries. Further, if one of the binary components is a BH, then not all of the GW power travels to spatial infinity. Instead, some of the GWs travel through the horizon of the BH, increasing its mass and spin throughout the evolution of the binary.  With these considerations, we can postulate that the generic PN formalism can be extended to include such effects by writing
\begin{equation}
\vec{A} = \vec{A}_{\rm PP} + \vec{A}_{\rm Spin} + \vec{A}_{\rm FS} + \vec{A}_{\rm H} + \vec{A}_{\rm ppE}\,,
\end{equation}
where $\vec{A} \in (\vec{P}, \vec{R}, \vec{\cal{V}}, \vec{\cal{D}})$. In the above, $\vec{A}_{\rm PP}$ represents the point particle terms, computed here to 3PN order, $\vec{A}_{\rm Spin}$ are the corrections generated by the spins of the compact objects, $\vec{A}_{\rm FS}$ are generated by finite size effects of NSs, and $\vec{A}_{\rm H}$ incorporates the corrections from the GWs fluxes through BH horizons. The final term, $\vec{A}_{\rm ppE}$ represents corrections due to modified theories of gravity, which have already been considered in~\cite{Loutrel:2014vja}.

One important question to address in the future concerns the most appropriate equations one should use to obtain the numerical evolution of highly elliptic systems under radiation reaction. In this work, we have used the orbit averaged equations for $\langle \dot{e}_{t} \rangle$ and $\langle \dot{x} \rangle$. These equations are applicable when the GW emission is smeared over the entire orbit and changes to the orbital elements are small on the timescale of one orbit, as  is the case in quasi-circular inspirals. However, for the highly elliptic binaries considered here, the GW emission is concentrated at pericenter passage, and changes to the orbital elements happen on timescales significantly shorter than the orbital period. The evolution of such binaries will resemble a set of discrete steps from one orbit to the next. Furthermore, it can be shown that when expressed in terms of variables that are finite in the parabolic limit, the orbit averaged fluxes of energy and angular momentum vanish for parabolic orbits. There is, of course, nothing special about the parabolic limit, and binaries on parabolic orbits will still emit GWs, which suggests a break down of the orbit-averaged formalism in this limit. Since the orbit averaged equations are currently used prolifically in the literature, it is important to determine how big of a deviation in observables is generated by considering evolutions with and without orbit-averaging in the radiation-reaction force, and what set of systems in $(f_{\rm GW}, e_{t})$ space are affected by this deviation. Such a study is currently underway~\cite{Loutrel-avg}.

Another avenue for future research is to consider how the 3PN order burst model aids in detecting highly elliptic binaries. In such a study, one would inject a waveform generated by numerically evolving the binary under radiation reaction into a simulated LIGO data stream. One could then perform an analysis to study whether the prior, specifically the burst model, is sufficient to achieve detection of such a signal given a particular noise model. One could also investigate the nature of posterior probability densities of recovered parameters and determine if such a search is accurate enough to perform parameter estimation on actual signals. With such a study completed, a follow up study could be conducted to investigate the search strategy's ability to estimate deviations from the current model, such as those from modified theories of gravity and to constrain the coupling constants of such theories. Such studies will be crucial for understanding our ability to detect and perform important astrophysics with eccentric GW signals.

\section*{Acknowledgements}
We would like to thank Frans Pretorius for several useful discussions. N.Y. acknowledges support from the NSF CAREER Grant PHY-1250636. N. L. acknowledges support from the NSF EAPSI Fellowship Award No. 1614203.
\appendix

\section{PN Recursion Relations}
\label{recursion}
When computing the burst model to arbitrary PN order, we are often faced will expressions of the form
\begin{equation}
\left(1 + \sum_{n=1}^{\infty} A_{n} x^{n}\right)^{-m}\,,
\end{equation}
which need to be perturbatively expanded about $x \ll 1$. In our PN burst model, $v_{p}$ takes the place of $x$. The expansion of the above expression can be easily computed term by term, but may not be expressible in terms of an arbitrary sum. We instead define the coefficients as
\begin{equation}
\sum_{n=1}^{\infty} A_{n}^{(-m)} x^{n} \equiv \left(1 + \sum_{n=1}^{\infty} A_{n} x^{n}\right)^{-m} - 1\,,
\end{equation}
where it is understood that we are working perturbatively in $x$. If one computes the Taylor expansion and calculates the coefficients, one finds
\begin{align}
\label{taylor}
A_{n}^{(-m)} &= -m A_{n} + \frac{m(m+1)}{2!} B_{n} - \frac{m(m+1)(m+2)}{3!} C_{n}
\nn \\
& + \frac{m(m+1)(m+2)(m+3)}{4!} D_{n} - ...\,,
\end{align}
where the coefficients $(B_{n}, C_{n}, D_{n})$ are given by
\begin{align}
\label{Bn}
B_{n} &= \sum_{q=1}^{n-1} A_{q} A_{n-q}\,,
\\
\label{Cn}
C_{n} &= \sum_{q=1}^{n-1} A_{q} B_{n-q}\,,
\\
\label{Dn}
D_{n} &= \sum_{q=1}^{n-1} B_{q} B_{n-q}\,.
\end{align}
We have stopped the expansion at fourth order in $x$, but in principle there will also be fifth order terms with coefficients $E_{n}$, sixth order terms with coefficients $F_{n}$, etc., where ($E_{n}$, $F_{n}$, ...)  are expressible in terms of the coefficients above. Thus, while the right-hand-side of Eq.~\eqref{taylor} is not expressible as a closed sum, one can recursively build the coefficients to arbitrary order.

As an example of how this works, consider the case where $m=1$ and we truncate at third order, i.e. the series only has coefficients $(A_{1}, A_{2}, A_{3})$. So, the first order term ($n=1$) is then
\begin{equation}
A_{1}^{(-1)} = - A_{1} + B_{1} - C_{1} + ...
\end{equation}
However, from Eq.~\eqref{Bn}
\begin{equation}
B_{1} = \sum_{q=1}^{0} A_{q} A_{1-q} = 0
\end{equation}
Likewise, all higher order terms, i.e. $(C_{1}, D_{1}, ...)$, will vanish, and thus at first order $A_{1}^{(-1)} = - A_{1}$ as one would expect from a first order Taylor expansion. At second order $n=2$,
\begin{equation}
A_{2}^{(-1)} = - A_{2} + B_{2} - C_{2} + ...
\end{equation}
However, now the $B$ coefficient doesn't vanish, but the $C$ coefficient (and thus all higher order coefficients) does vanish.
\begin{align}
B_{2} &= \sum_{q=1}^{1} A_{q} A_{2-q} = A_{1}^{2}
\\
C_{2} &= \sum_{q=1}^{1} A_{q} B_{2-q} = A_{1} B_{1} = 0
\end{align}
So, the second order coefficient is $A_{2}^{(-1)} = - A_{2} + A_{1}^{2}$. Similarly, a calculation of the third order coefficient gives $A_{3}^{(-1)} = -A_{3} + 2 A_{1} A_{2} - A_{1}^{3}$. Since we are working perturbatively to third order in this example, we now truncate the series and all higher order coefficients will be zero.

The expression in Eq.~\eqref{taylor} can be written in a more compact form using the notation of Section~\ref{PN}. Using the vector convolution $\circ$, we may write
\begin{equation}
A_{n}^{(-m)} = \sum_{k=1}^{n} \binom {-m} {k-1} \left[(\vec{A} \circ)^{k-1} \vec{A}\right]_{n}\,.
\end{equation}
%
\section{PN Vector Fields}
\label{fields}
We here give explicit expressions for the PN amplitude vector fields $(P_{k}, R_{k}, V_{k}, E_{k})$ to 3PN order. We will provide an example calculation at 1PN order to show how to obtain these functions. At higher PN order, we will simply give the results of the calculation.

\subsection{1PN Amplitude Vector Fields}
We begin by calculating the 1PN corrections to the factors $(P_{2}, R_{2}, V_{2}, E_{2})$. This naturally amounts to finding the 1PN corrections to the orbital period, pericenter distance, and rates of change of pericenter velocity and orbital eccentricity. Because these are not quantities that are typically computed in the PN literature, we will show explicitly how to calculate them from other known quantities. 

Consider first the pericenter velocity and the orbital eccentricity as functions of the orbital energy and angular momentum. Typically, PN quantities are written in terms of the reduced orbital energy $\varepsilon = - 2 E/\mu$ and the dimensionless orbital angular momentum $j = - 2 E L^{2}/ \mu^{3} M^{2}$, where $\mu$ is the reduced mass of the system. The reason for this is that $\varepsilon$ and $j$ do not depend on the coordinate system that one chooses to do calculations in, i.e. they are coordinate invariant quantities. However, we have formulated our burst model in terms of the pericenter velocity and orbital eccentricity, so we need to determine the mappings $(\varepsilon, j) \rightarrow (v_{p}, e)$.

We start from the equations of motion that govern the QK representation, which to 1PN order are~\cite{Blanchet:2013haa}
\begin{align}
\label{r-eqn}
r &= a_{r} \left(1 - e_{r} {\rm cos} \; u\right)\,,
\\
\label{t-eqn}
\ell &= u - e_{t} {\rm sin} \; u\,,
\\
\label{phi-eqn}
\phi - \phi_{0} &= 2 K \; \arctan \left[ \left(\frac{1 + e_{\phi}}{1 - e_{\phi}}\right)^{1/2} {\rm tan} \left(\frac{u}{2}\right)\right]\,,
\end{align}
The first of these is the radial equation of the elliptical orbit written in terms of the eccentric anomaly $u$, where $a_{r}$ is the semi-major axis of the ellipse and $e_{r}$ is the radial eccentricity. The second equation is Kepler's equation, which relates the eccentric anomaly to the mean anomaly through the time eccentricity $e_{t}$. This equation is itself a direct measure of time since $\ell = n (t - t_{p})$ where $n=2\pi/P$ is the mean motion and $t_{p}$ is the time of pericenter passage. Finally, the last equation is the azimuthal equation of the orbit, which relates the orbital phase $\phi$ to the eccentric anomaly through the azimuthal eccentricity $e_{\phi}$ and the advance of periastron per orbit $K$. In Newtonian gravity, $K=1$ and we recover the Newtonian equations of motion, but at 1PN order $K=1 + {\cal{O}}(\varepsilon)$. Thus a primary difference in the azimuthal motion at 1PN order is the inclusion of the precession of periastron. 

Another difference between Newtonian and 1PN orbits in the QK parametrization is the need for multiple "eccentricities." The latter are not actually separate physical quantities, but instead are a clever notational trick that allows the equations of motion to take the same functional form as their Newtonian analogs\footnote{Note that while this is true to 1PN order, the same cannot be said at higher PN order. At 2PN order and beyond, Eqs~\eqref{t-eqn} and~\eqref{phi-eqn} pick up higher harmonics of the anomalies.}. In reality, these eccentricities are related to the Newtonian expression for the orbital eccentricity, which is just $e = \sqrt{1 - j}$, through~\cite{Blanchet:2013haa} 
\begin{align}
\label{er-exp}
e_{r} &= \sqrt{1 - j} + \frac{\varepsilon}{8 \sqrt{1 - j}} \left[24 - 4 \eta + 5 j (-3 + \eta)\right]\,,
\\
\label{et-exp}
e_{t} &= \sqrt{1 - j} + \frac{\varepsilon}{8 \sqrt{1 - j}} \left[-8 + 8 \eta + j (17 - 7 \eta)\right]\,,
\\
\label{ephi-exp}
e_{\phi} &= \sqrt{1 - j} + \frac{\varepsilon}{8 \sqrt{1 - j}} \left[24 + j (-15 + \eta)\right]\,.
\end{align}
In this paper, and in this Appendix, we choose to parameterize the motion in terms of one of these eccentricities, specifically $e_{t}$, and we define $\delta e_{t} = 1 - e_{t}$. This choice is arbitrary: one could easily construct the burst model in terms of $e_{r}$ or $e_{\phi}$ or the Newtonian $e$. With this choice, we now have one of the equations we need to determine $\varepsilon(v_{p}, e_{t})$ and $j(v_{p}, e_{t})$, specifically Eq.~\eqref{et-exp}.

Let us now focus on the pericenter velocity. Since pericenter is the minimum turning point of the orbit ($\dot{r} = 0$), we may write $v_{p} = r_{p} \dot{\phi}(r=r_{p})$. From Eq.~\eqref{r-eqn}, we see that $r$ is at a minimum when $u=0$, specifically $r(u=0) = r_{p} = a_{r} (1 - e_{r})$. Hence we are left with determining $\dot{\phi}(u)$. Taking the time derivative of Eq.~\eqref{phi-eqn}, we find
\begin{align}
\dot{\phi} &= \frac{K \beta \left[1 + {\rm tan}^{2}\left(\frac{u}{2}\right)\right]}{1 + \beta^{2} {\rm tan}^{2}\left(\frac{u}{2}\right)} \dot{u}\,,
\end{align}
where $\beta = \sqrt{(1 + e_{\phi})/(1 - e_{\phi})}$. To find the expression for $\dot{u}$, we take the time derivative of Eq.~\eqref{t-eqn} and solve for $\dot{u}$,
\begin{align}
\dot{u} &= \frac{n}{1 - e_{t} {\rm cos} \; u}\,.
\end{align}
We can now put all of this together to find $v_{p}$ to 1PN order:
\begin{equation}
v_{p} = a_{r} n K \beta \frac{1 - e_{r}}{1 - e_{t}}\,.
\end{equation}

This expression for the pericenter velocity is exact in the sense that we have not performed a PN expansion in $\varepsilon$ yet. To obtain $v_{p}(\varepsilon,j)$, we use Eqs.~\eqref{er-exp}-\eqref{ephi-exp} with~\cite{Blanchet:2013haa}
\begin{align}
\label{eq:n}
n &= \frac{\varepsilon^{3/2}}{M} \left[1 + \frac{\varepsilon}{8} (-15 + \eta)\right]\,,
\\
K &= 1 + \frac{3 \varepsilon}{j}\,,
\\
a_{r} &= \frac{M}{\varepsilon} \left[1 + \frac{\varepsilon}{4} \left(-7 + \eta\right)\right]\,,
\end{align}
and expand in $\varepsilon$ to find
\begin{align}
\label{eq:vp}
v_{p} &= \frac{\varepsilon^{1/2} j^{7/2}}{\left(1 - \sqrt{1 - j}\right)^{4} \left(1 + \sqrt{1 - j}\right)^{3}}  - \frac{\varepsilon^{3/2} j^{5/2}}{(1 - j) \left(1 - \sqrt{1 - j}\right)^{4} \left(1 + \sqrt{1 - j}\right)^{3}}
\nn \\
& \times \left\{8 - 12 \eta + j(-11 + 21 \eta) + j^{2} (3 - 9 \eta) + \sqrt{1-j} \left[8 - 12 \eta + j (-17 + 11 \eta)\right]\right\}\,.
\end{align}

The last step before deriving the functions $(P_{2}, R_{2})$ is to invert Eqs.~\eqref{et-exp} and~\eqref{eq:vp} to find $\varepsilon(v_{p},e_{t})$ and $j(v_{p},e_{t})$. This can be done very easily order by order in $\varepsilon$ (or $v_{p}$) to find
\begin{align}
\varepsilon &= v_{p}^{2} \frac{1 - e_{t}}{1 + e_{t}} \left[1 - \frac{v_{p}^{2}}{4} \;\; \frac{-5 + 3 \eta + 4 \eta e_{t} + e_{t}^{2} \left(-3 + 9 \eta\right)}{(1 + e_{t})^{2}} \right]\,,
\\
j &= \left(1 - e_{t}^{2}\right) \left[1 + \frac{v_{p}^{2}}{4} \; \frac{9 + \eta + e_{t}^{2} \left(-17 + 7 \eta\right)}{(1 + e_{t})^{2}}\right]\,.
\end{align}

We can now construct the functions $P_{2}$ and $R_{2}$, which recall we define via 
\begin{align}
P &= P^{\rm N} \left(1 + v^{2}_{p} \; P_{2} \right)\,,
\\
\label{rp-1PN}
r_{p} &= r_{p}^{\rm N} \left(1 + v_{p}^{2} \; R_{2} \right)\,.
\end{align} 
The orbital period in the QK representation 
is given by $P = 2 \pi/n(\varepsilon,j)$, with $n(\varepsilon,j)$ given in Eq.~\eqref{eq:n}. Inserting our expressions for $\varepsilon(v_{p}, e_{t})$ and $j(v_{p}, e_{t})$ and expanding about $v_{p} \ll 1$, while keeping terms of relative ${\cal{O}}(v_{p}^{2})$, we find
\begin{equation}
P_{2}(e_{t}, \eta) = \frac{2 \eta + 3 \eta e_{t} + e_{t}^{2} (-6 + 7 \eta)}{2 (1 + e_{t})^{2}}\,.
\end{equation}
We may follow the same procedure for the pericenter distance, using $r_{p} = a_{r}(\varepsilon, j) \left[1 - e_{r}(\varepsilon, j)\right]$; we find
\begin{equation}
\label{R2}
R_{2}(e_{t}, \eta) = \frac{-6 + 2 \eta + e_{t} (-8 + 5 \eta) + e_{t}^{2} (-6 + 7 \eta)}{2 (1 + e_{t})^{2}}\,.
\end{equation}

The functions $(V_{2}, E_{2})$ require a bit more work, since we need to compute the rates of change of pericenter velocity and time eccentricity. We may do this via the chain rule:
\begin{align}
\label{chain-vp}
\langle \dot{v}_{p} \rangle &= \frac{\partial v_{p}(\varepsilon, h)}{\partial \varepsilon} \left(\frac{-2 \langle \dot{E} \rangle}{\mu}\right) + \frac{\partial v_{p}(\varepsilon, h)}{\partial h} \left(\frac{\langle \dot{L} \rangle}{M}\right)\,,
\\
\label{chain-et}
\langle \dot{e}_{t} \rangle &= \frac{\partial e_{t}(\varepsilon, h)}{\partial \varepsilon} \left(\frac{-2 \langle \dot{E} \rangle}{\mu}\right) + \frac{\partial e_{t}(\varepsilon, h)}{\partial h} \left(\frac{\langle \dot{L} \rangle}{M}\right)\,,
\end{align}
where $h = L/M$ is the reduced angular momentum, $\dot{E}$ is the energy flux, and $\dot{L}$ is the angular momentum flux. The reason for replacing $j$ with $h$ in the expressions $\left[v_{p}(\varepsilon,j), e_{t}(\varepsilon,j)\right]$ is that $j$ has a factor of the reduced energy hidden in it, which needlessly obfuscates taking partial derivatives. Furthermore, the expression for $\dot{h}$ can be computed directly and easily from $\dot{L}$. The averaged energy and angular momentum fluxes are given to 1PN order by~\cite{Arun:2007sg,Arun:2009mc}
\begin{align}
\langle\dot{E}\rangle &= - \frac{32}{5} \eta^{2} x^{5} \left({\cal{I}}_{\rm N} + x \; {\cal{I}}_{\rm 1PN}\right)
\\
\langle\dot{L}\rangle &= - \frac{4}{5} \eta^{2} M x^{7/2} \left({\cal{G}}_{\rm N} + x \; {\cal{G}}_{\rm 1PN}\right)
\end{align}
with the enhancement factors
\begin{align}
{\cal{I}}_{\rm N} &= \frac{1+\frac{73}{24} e_{t}^{2}+\frac{37}{96} e_{t}^{4}}{(1 - e_{t}^{2})^{7/2}}\,,
\\
{\cal{I}}_{\rm 1PN} &= \frac{1}{(1 - e_{t}^{2})^{9/2}} \left[-\frac{1247}{336}-\frac{35}{12} \eta + e_{t}^{2} \left(\frac{10475}{672} - \frac{1081}{36}\eta\right)
 + e_{t}^{4} \left(\frac{10043}{384} - \frac{311}{12} \eta \right)
\right.
\nn \\
&\left.
 + e_{t}^{6} \left(\frac{2179}{1792} - \frac{851}{576} \eta \right)\right]\,,
\end{align}
and
\begin{align}
{\cal{G}}_{\rm N} &= \frac{8 + 7e_{t}^{2}}{(1 - e_{t}^{2})^{2}}\,,
\\
{\cal{G}}_{\rm 1PN} &= \frac{1}{(1 - e_{t}^{2})^{3}} \left[-\frac{1247}{42} - \frac{70}{3}\eta + e_{t}^{2} \left(\frac{3019}{42} - \frac{335}{3} \eta \right) + e_{t}^{4} \left(\frac{8399}{336} - \frac{275}{12} \eta \right)\right]\,.
\end{align}
In the above, the PN expansion parameter $x = (M \Omega)^{2/3}$, with $\Omega$ the orbital frequency, which is related to the reduced orbital energy by~\cite{Blanchet:2013haa}
\begin{align}
x &= \varepsilon \left[1 + \varepsilon \left(-\frac{5}{4} + \frac{\eta}{12} + \frac{2}{j}\right)\right]\,.
\end{align}

The goal at this stage should be clear: we desire to write $\langle \dot{v}_{p} \rangle $ and $\langle \dot{e}_{t} \rangle$ in terms of $v_{p}$ and $e_{t}$ instead of $x$ and $e_{t}$. The calculation is rather lengthy, but straightforward, so we will simply detail the main steps here. The first step is to replace $x$ with $\varepsilon$ in the expressions for $\langle\dot{E}\rangle$ and $\langle\dot{L}\rangle$, working perturbatively in $\varepsilon$. We then use the chain rule combined with our new expressions for the fluxes in Eqs.~\eqref{chain-vp} and~\eqref{chain-et}. Finally we replace any instances of $(\varepsilon, h)$ with $(v_{p}, e_{t})$ and expand in powers of $v_{p}$. The end result is
\begin{align}
\langle \dot{v}_{p} \rangle &= \langle \dot{v}_{p}^{\rm N} \rangle (v_{p}, e_{t})  \left[1 + v_{p}^{2} \; V_{2}(e_{t}, \eta)\right]\,,
\\
\langle \dot{e}_{t} \rangle &= \langle \dot{e}^{\rm N} \rangle (v_{p}, e_{t}) \left[1 + v_{p}^{2} \; E_{2}(e_{t}, \eta) \right]\,,
\end{align}
where $\langle \dot{v}_{p}^{\rm N} \rangle$ and $\langle \dot{e}^{\rm N} \rangle$ are given in Eqs.~\eqref{vp-dot-N} and~\eqref{e-dot-N} with the replacement $e \rightarrow e_{t}$, and the PN functions are
\begin{align}
V_{2} (e_{t}, \eta) &= \frac{1}{(1 + e_{t})^{2} \left(1 - \frac{13}{6} e_{t} + \frac{7}{8} e_{t}^{2} - \frac{37}{96} e_{t}^{3}\right)} \left[- \frac{19}{4} \eta + \frac{1273}{336} + \left(\frac{319}{24} \eta - \frac{3887}{224}\right) e_{t} 
\right.
\nn \\
&\left.
+ \left(-\frac{421}{16} \eta + \frac{1507}{42}\right) e_{t}^{2} + \left(\frac{1711}{48} \eta - \frac{29605}{672}\right) e_{t}^{3} + \left(-\frac{737}{64} \eta + \frac{35923}{2688}\right) e_{t}^{4} 
\right.
\nn \\
&\left.
+ \left(\frac{407}{96} \eta - \frac{24149}{5376}\right) e_{t}^{5} \right]\,,
\\
E_{2} (e_{t}, \eta) &= \frac{1}{(1+ e_{t})^{2} \left(1 + \frac{121}{304} e_{t}^{2}\right)} \left[ \frac{543}{76} \eta - \frac{14207}{2128} + 4 e_{t} \eta + \left(\frac{14073}{608} \eta - \frac{53717}{2128}\right) e_{t}^{2} + \frac{121}{76} e_{t}^{3} \eta 
\right.
\nn \\
&\left.
+ \left( \frac{421}{76} \eta - \frac{95995}{17024}\right) e_{t}^{4}\right]\,.
\end{align}
This completes the construction of the 1PN correction functions $(P_{2}, R_{2}, V_{2}, E_{2})$.

\subsection{Amplitude Vector Fields to 3PN Order}

We now provide the components of the amplitude vector fields $(\vec{P}, \vec{R}, \vec{V}, \vec{E})$ to 3PN order. The derivation of the components follows the exact same procedure as the previous section, so we will simply list the components, where recall we work in ADM coordinates. For the orbital period, the non-zero components are
\begin{align}
P_{4}(e_{t}, \eta; v_{p}) &= \frac{3 (1 - e_{t}) (1 + 3 e_{t} - e_{t}^{2}) (5 - 2 \eta)}{2 (1 + e_{t})^{3} \sqrt{1 - e_{t}^{2}}} + \frac{1}{8 (1 + e_{t})^{4}} \left[19\eta - 6 + e_{t} \left(-23\eta^{2} + 192\eta - 258\right) 
\right.
\nn \\
&\left.
+ e_{t}^{2} \left(-75\eta^{2} + 242\eta - 186\right) + e_{t}^{3} \left(-37\eta^{2} + 42\eta\right) + e_{t}^{4} \left(-21\eta^{2} - 21\eta\right)\right]\,,
\\
P_{6}(e_{t}, \eta; v_{p}) &= \frac{(1 - e_{t})}{192 (1 + e_{t})^{5} \sqrt{1 - e_{t}^{2}}} \left[123\pi^{2}\eta + 1152\eta^{2} - 13664\eta + 8640 
\right.
\nn \\
&\left.
+ e_{t} \left(369\pi^{2}\eta + 4032\eta^{2} - 42432\eta + 25920\right) 
\right.
\nn \\
&\left.
+ e_{t}^{2} \left(-123\pi^{2}\eta - 2016\eta^{2} + 18560\eta - 12960\right) 
\right.
\nn \\
&\left.
+ e_{t}^{3} \left(8064\eta^{2} - 25776\eta + 21600\right) + e_{t}^{4} \left(-1728\eta^{2} + 5040\eta - 4320\right)\right]
\nn \\
&+\frac{1}{192 e_{t} (1 + e_{t})^{6}} \left[-6\eta + e_{t} \left(1116\pi^{2}\eta - 1536\eta^{2} - 20088\eta + 2976\right) 
\right.
\nn \\
&\left.
+ e_{t}^{2} \left(783\pi^{2}\eta + 684\eta^{3} - 16260\eta^{2} + 51264\eta - 38304\right) 
\right.
\nn \\
&\left.
+ e_{t}^{3} \left(396\pi^{2}\eta + 3684\eta^{3} - 31572\eta^{2} + 56208\eta - 34128\right) 
\right.
\nn \\
&\left.
+ e_{t}^{4} \left(9\pi^{2}\eta + 7320\eta^{3} - 46056\eta^{2} + 77106\eta - 45936\right) 
\right.
\nn \\
&\left.
+ e_{t}^{5} \left(10248\eta^{3} - 31488\eta^{2} + 36708\eta - 20016\right) 
\right.
\nn \\
&\left.
+ e_{t}^{6} \left(4332\eta^{3} - 5436\eta^{2} + 2844\eta\right) + e_{t}^{7} \left(1764\eta^{3} + 156\eta^{2} - 228\eta\right)\right]\,.
\end{align}

For the pericenter distance, the non-zero components are
\begin{align}
\label{R4}
R_{4}(e_{t}, \eta; v_{p}) &= \frac{3 e_{t} (1 - e_{t})^{2} (5 - 2 \eta)}{2 (1 + e_{t})^{3} \sqrt{1 - e_{t}^{2}}} + \frac{1}{8 (1 + e_{t})^{4}} \left[43\eta - 30 + e_{t} \left(-21\eta^{2} + 184\eta - 206\right) 
\right.
\nn \\
&\left.
+ e_{t}^{2} \left(-63\eta^{2} + 184\eta - 140\right) + e_{t}^{3} \left(-31\eta^{2} + 2\eta\right) + e_{t}^{4} \left(-21\eta^{2} - 21\eta\right)\right]\,,
\\
\label{R6}
R_{6}(e_{t}, \eta; v_{p}) &= \frac{e_{t} (1 - e_{t})^{2}}{192 (1 + e_{t})^{5} \sqrt{1 - e_{t}^{2}}} \left[123\pi^{2}\eta + 864\eta^{2} - 12368\eta + 7200 
\right.
\nn \\
&\left.
+ e_{t} \left(-2880\eta^{2} + 10656\eta - 8640\right) + e_{t}^{2} \left(1728\eta^{2} - 5040\eta + 4320\right)\right] 
\nn \\
&+ \frac{1}{192 e_{t} (1 + e_{t})^{6}} \left[-4\eta + e_{t} \left(501\pi^{2}\eta - 2112\eta^{2} - 1804\eta - 2640\right) 
\right.
\nn \\
&\left.
+ e_{t}^{2} \left(279\pi^{2}\eta + 660\eta^{3} - 14724\eta^{2} + 48952\eta - 30288\right) 
\right.
\nn \\
&\left.
+ e_{t}^{3} \left(267\pi^{2}\eta + 3300\eta^{3} - 29508\eta^{2} + 50888\eta - 27792\right) 
\right.
\nn \\
&\left.
+ e_{t}^{4} \left(9\pi^{2}\eta + 6648 \eta^{3} - 40392\eta^{2} + 68584\eta - 39696\right) 
\right.
\nn \\
&\left.
+ e_{t}^{5} \left(9336\eta^{3} - 26520\eta^{2} + 31704\eta - 16704\right) 
\right.
\nn \\
&\left.
+ e_{t}^{6} \left(4020\eta^{3} - 3636\eta^{2} + 2292\eta\right) + e_{t}^{7} \left(1764\eta^{3} + 156\eta^{2} - 228\eta\right)\right]\,.
\end{align}

For $\langle \dot{e}_{t} \rangle$, we have
\begin{align}
E_{3}(e_{t}, \eta; v_{p}) &= \frac{985 \pi (1 - e_{t})^{3} (1 - e_{t}^{2}) \varphi_{e}(e_{t})}{152 (1 + \frac{121}{304} e_{t}^{2})}\,,
\\
E_{4}(e_{t}, \eta; v_{p}) &= \frac{1}{(1 + e_{t})^{4} (1 + \frac{121}{304} e_{t}^{2})} \left[\frac{1035}{38}\eta^{2} - \frac{45819}{608}\eta + \frac{1365463}{38304} + e_{t} \left(\frac{3399}{76}\eta^{2} - \frac{207227}{2128}\eta + 86\right) 
\right.
\nn \\
&\left.
+ e_{t}^{2} \left(\frac{599243}{2432}\eta^{2} - \frac{9115835}{17024}\eta + \frac{13816637}{51072}\right) + e_{t}^{3} \left(\frac{82879}{608}\eta^{2} - \frac{361097}{2128}\eta + \frac{5203}{152}\right) 
\right.
\nn \\
&\left.
+ e_{t}^{4} \left(\frac{97141}{304}\eta^{2} - \frac{21541341}{34048}\eta + \frac{108773587}{306432}\right) + e_{t}^{5} \left(\frac{10477}{304}\eta^{2} - \frac{601943}{17024}\eta\right) 
\right.
\nn \\
&\left.
+ e_{t}^{6} \left(\frac{15933}{304}\eta^{2} - \frac{97941}{1064}\eta + \frac{3284783}{68096}\right)\right] + \frac{\sqrt{1 - e_{t}^{2}}}{(1 + e_{t})^{4} (1 + \frac{121}{304} e_{t}^{2})} \left[-\frac{15}{19}\eta + \frac{75}{38} 
\right.
\nn \\
&\left.
+ e_{t} \left(24\eta - 60\right) + e_{t}^{2} \left(-\frac{8427}{304}\eta + \frac{42135}{608}\right) + e_{t}^{3} \left(\frac{363}{38}\eta - \frac{1815}{76}\right) 
\right.
\nn \\
&\left.
+ e_{t}^{4} \left(-\frac{1533}{304}\eta + \frac{7665}{608}\right)\right]\,,
\\
E_{5}(e_{t}, \eta; v_{p}) &= \frac{\pi (1 - e_{t})^{3} (1 - e_{t}^{2}) \varphi_{e}(e_{t})}{(1 + e_{t})^{2} (1 + \frac{121}{304} e_{t}^{2})} \left[-\frac{10835}{456}\eta + \frac{10835}{152} - \frac{10835}{304} e_{t} \eta + e_{t}^{2} \left(-\frac{75845}{912}\eta + \frac{10835}{152} \right)\right] 
\nn \\
&- \frac{\pi (1 - e_{t})^{5}}{(1 + \frac{121}{304} e_{t}^{2})} \left[\frac{55691}{4256} \psi_{e}(e_{t}) + \frac{19067 \eta}{399} \zeta_{e}(e_{t})\right]\,,
\end{align}
\begin{align}
E_{6}(e_{t}, \eta; v_{p}) &= \frac{(1 - e_{t})^{5} \sqrt{1 - e_{t}^{2}} \left[-\frac{89789209}{1117200} + \frac{4601}{105} {\rm ln}(2) - \frac{234009}{5320} {\rm ln}(3)\right]}{(1 + e_{t})(1 + \frac{121}{304} e_{t}^{2})} \kappa_{e}(e_{t}) 
\nn \\
& - \frac{1}{e_{t} (1 - e_{t}) (1 + e_{t})^{7} (1 + \frac{121}{304} e_{t}^{2})} \left\{\sum_{k=0}^{11} {\cal{P}}_{e}^{(k)}(\eta) e_{t}^{k}  + {\rm ln}\left[\frac{(1-e_{t})^{3/2} (1 + e_{t})^{1/2} v_{p}}{1 + \sqrt{1 - e_{t}^{2}}}\right]
\right.
\nn \\
&\left.
\times \left(-\frac{61311}{21280} e_{t}^{9} - \frac{880931}{12768} e_{t}^{7} - \frac{82283}{1064} e_{t}^{5} + \frac{71797}{665} e_{t}^{3} + \frac{82283}{1995} e_{t}\right)\right\}
\nn \\
&- \frac{\sqrt{1 - e_{t}^{2}}}{(1 - e_{t}) (1 + e_{t})^{7} (1 + \sqrt{1 - e_{t}^{2}}) (1 + \frac{121}{304} e_{t}^{2})} \left[\frac{91271}{59850} + \sum_{k=2}^{10} {\cal{S}}_{e}^{(k)}(\eta) e_{t}^{k} \right]
\end{align}
where $\gamma_{E} = 0.577...$ is the Euler constant and the tail enhancement factors $[\varphi_{e}(e_{t}), \psi_{e}(e_{t}), \zeta_{e}(e_{t}), \kappa_{e}(e_{t})]$ are defined in~\cite{Arun:2009mc}. The functions $[{\cal{P}}_{e}^{(k)}(\eta), {\cal{S}}_{e}^{(k)}(\eta)]$ are polynomials in $\eta$, specifically
\begin{align}
{\cal{P}}_{e}^{(0)}(\eta) &= -\frac{\eta}{12}\,,
\\
{\cal{P}}_{e}^{(1)}(\eta) &= \frac{370599031}{459648}\eta - \frac{24177239}{51072}\eta^{2} + \frac{46101}{608}\eta^{3} + \frac{82283}{1995} \gamma_{E} - \frac{72137736667}{126403200} - \frac{769}{57}\pi^{2}
\nn \\
&+ \frac{82283}{665} {\rm ln}(2) - \frac{2011}{1216}\pi^{2} \eta \,,
\\
{\cal{P}}_{e}^{(2)}(\eta) &= -\frac{9135439}{8512}\eta^{2} + \frac{23802845}{25536}\eta - \frac{1655487}{4256} + \frac{77915}{304}\eta^{3} + 16\pi^{2} \eta\,,
\\
{\cal{P}}_{e}^{(3)}(\eta) &= -\frac{595512727}{102144}\eta^{2} + \frac{6892277}{4864}\eta^{3} + \frac{71797}{665}\gamma_{E} - \frac{6899473429}{1663200} - \frac{671}{19}\pi^{2} + \frac{28354579}{3591}\eta 
\nn \\
&+ \frac{215391}{665} {\rm ln}(2) - \frac{13525}{9728}\pi^{2} \eta\,,
\\
{\cal{P}}_{e}^{(4)}(\eta) &= -\frac{21014717}{4256} \eta^{2} - \frac{917485}{532} + \frac{137723689}{25536}\eta + \frac{4144843}{2432} \eta^{3} - \frac{278}{19}\pi^{2} \eta\,,
\\
{\cal{P}}_{e}^{(5)}(\eta) &= -\frac{3788393467}{408576}\eta^{2} - \frac{82283}{1064}\gamma_{E} + \frac{16291239}{4864}\eta^{3} + \frac{3845}{152}\pi^{2} + \frac{30463753231}{3677184}\eta
\nn \\
&- \frac{502325236657}{202245120} - \frac{246849}{1064} {\rm ln}(2) + \frac{619095}{38912}\pi^{2} \eta\,,
\\
{\cal{P}}_{e}^{(6)}(\eta) &= -\frac{16539923}{68096} \eta^{2} + \frac{2092715}{51072}\eta - \frac{20806067}{34048} + \frac{745669}{1216} \eta^{3} - \frac{1021}{304}\pi^{2} \eta\,,
\\
{\cal{P}}_{e}^{(7)}(\eta) &= \frac{3175811627}{408576} \eta^{2} - \frac{35817065233}{3677184}\eta - \frac{85979}{64} \eta^{3} - \frac{880931}{12768}\gamma_{E} + \frac{41165}{1824}\pi^{2}
\nn \\
&+ \frac{80176348783}{14981120} - \frac{880931}{4256} {\rm ln}(2) - \frac{745}{64}\pi^{2} \eta\,,
\\
{\cal{P}}_{e}^{(8)}(\eta) &= \frac{10498719}{4864} + \frac{88818089}{17024} \eta^{2} - \frac{4689915}{896}\eta - \frac{5134385}{2432} \eta^{3} + \frac{605}{304}\pi^{2} \eta\,,
\\
{\cal{P}}_{e}^{(9)}(\eta) &= -\frac{467431}{152} \eta^{3} - \frac{61311}{21280}\gamma_{E} - \frac{4101817087}{612864}\eta + \frac{573}{608}\pi^{2} + \frac{706474273}{102144} \eta^{2} 
\nn \\
&+ \frac{4808376554393}{2696601600} - \frac{183933}{21280} {\rm ln}(2) - \frac{47683}{38912}\pi^{2} \eta\,,
\\
{\cal{P}}_{e}^{(10)}(\eta) &= \frac{507315}{896} + \frac{10083793}{9728} \eta^{2} - \frac{4818031}{4256}\eta - \frac{281279}{608} \eta^{3}\,,
\\
{\cal{P}}_{e}^{(11)}(\eta) &= -\frac{64379}{152}\eta^{3} + \frac{348257375}{5136384} - \frac{18848765}{34048}\eta + \frac{538919}{608}\eta^{2}\,,
\\
{\cal{S}}_{e}^{(2)}(\eta) &= -\frac{697}{1216}\pi^{2} \eta - \frac{351}{76}\eta^{2} + \frac{7230223}{136800} + \frac{358327}{6384}\eta\,,
\\
{\cal{S}}_{e}^{(3)}(\eta) &= \frac{10245}{38}\eta^{2} + \frac{1831605}{2128} - \frac{4225037}{3192}\eta + \frac{41}{8}\pi^{2} \eta\,,
\\
{\cal{S}}_{e}^{(4)}(\eta) &= -\frac{103033}{19456}\pi^{2} \eta + \frac{41831239}{25536}\eta - \frac{185949}{608}\eta^{2} - \frac{139310891}{106400}\,,
\\
{\cal{S}}_{e}^{(5)}(\eta) &= \frac{129609}{304}\eta^{2} - \frac{6189835}{4256}\eta + \frac{1634085}{1064} - \frac{7503}{2432}\pi^{2} \eta\,,
\\
{\cal{S}}_{e}^{(6)}(\eta) &= \frac{50061}{9728}\pi^{2} \eta - \frac{261333}{608}\eta^{2} + \frac{895061}{896}\eta - \frac{6498062887}{7660800}\,,
\\
{\cal{S}}_{e}^{(7)}(\eta) &= -\frac{31159215}{17024} - \frac{152823}{304}\eta^{2} + \frac{53246143}{25536}\eta - \frac{4961}{2432}\pi^{2} \eta\,,
\\
{\cal{S}}_{e}^{(8)}(\eta) &= \frac{14063}{19456}\pi^{2} \eta + \frac{3537630247}{1915200} + \frac{387195}{608}\eta^{2} - \frac{59812661}{25536}\eta\,,
\\
{\cal{S}}_{e}^{(9)}(\eta) &= -\frac{29373}{152}\eta^{2} + \frac{5897721}{8512}\eta - \frac{507315}{896}\,,
\\
{\cal{S}}_{e}^{(10)}(\eta) &= \frac{8717115}{34048} - \frac{5974083}{17024}\eta + \frac{62895}{608}\eta^{2}\,.
\end{align}

Finally, for $\langle \dot{v}_{p} \rangle$, we have
\begin{align}
V_{3}(e_{t}, \eta; v_{p}) &= \frac{4 \pi (1 - e_{t})^{4} (1 + e_{t})^{2} \left[\tilde{\varphi}(e_{t}) - (1 - e_{t}) \sqrt{1 - e_{t}^{2}} \varphi(e_{t})\right]}{e_{t} \sqrt{1 - e_{t}^{2}} \left(1 - \frac{13}{6} e_{t} + \frac{7}{8} e_{t}^{2} - \frac{37}{96} e_{t}^{3}\right)}\,,
\\
V_{4}(e_{t}, \eta; v_{p}) &= \frac{3 (1 - e_{t}) \left(1 + 9 e_{t} - \frac{219}{8} e_{t}^{2} + \frac{1039}{48} e_{t}^{3} - \frac{251}{32} e_{t}^{4} + \frac{37}{24} e_{t}^{5}\right) (-5 + 2 \eta)}{2 (1 + e_{t})^{3} \sqrt{1 - e_{t}^{2}} \left(1 - \frac{13}{6} e_{t} + \frac{7}{8} e_{t}^{2} - \frac{37}{96} e_{t}^{3}\right)} 
\nn \\
&+ \frac{1}{(1 + e_{t})^{4} \left(1 - \frac{13}{6} e_{t} + \frac{7}{8} e_{t}^{2} - \frac{37}{96} e_{t}^{3}\right)} \left[\frac{27}{2}\eta^{2} - \frac{74243}{2016}\eta + \frac{411671}{18144} 
\right.
\nn \\
&\left.
+ e_{t} \left(-\frac{275}{8}\eta^{2} + \frac{119705}{1008}\eta - \frac{477359}{36288}\right) + e_{t}^{2} \left(\frac{34627}{192}\eta^{2} - \frac{246245}{672}\eta + \frac{1200901}{12096}\right) 
\right.
\nn \\
&\left.
+ e_{t}^{3} \left(-\frac{18457}{64}\eta^{2} + \frac{4580099}{5376}\eta - \frac{24721289}{48384}\right) + e_{t}^{4} \left(\frac{272143}{768}\eta^{2} - \frac{1484453}{1792}\eta + \frac{26632505}{48384}\right) 
\right.
\nn \\
&\left.
+ e_{t}^{5} \left(-\frac{273941}{768}\eta^{2} + \frac{363297}{448}\eta - \frac{16657127}{32256}\right) + e_{t}^{6} \left(\frac{77971}{768}\eta^{2} - \frac{2128543}{10752}\eta + \frac{1922899}{16128}\right) 
\right.
\nn \\
&\left.
+ e_{t}^{7} \left(-\frac{3145}{96}\eta^{2} + \frac{108099}{1792}\eta - \frac{2162753}{64512}\right)\right]\,,
\\
V_{5}(e_{t}, \eta; v_{p}) &= \frac{30 \pi (1 - e_{t})^{4} \left[1 - \frac{8}{45} \eta + e_{t} \left(-\frac{1}{3}\eta - \frac{4}{15}\right) + e_{t}^{2} \left(-\frac{79}{45}\eta + \frac{5}{3}\right)\right]}{e_{t} \sqrt{1 - e_{t}^{2}} \left(1 - \frac{13}{6} e_{t} + \frac{7}{8} e_{t}^{2} - \frac{37}{96} e_{t}^{3}\right)} \tilde{\varphi}(e_{t})
\nn \\
&- \frac{30 \pi (1 - e_{t})^{5} \left[1 - \frac{8}{45} \eta + e_{t} \left(-\frac{2}{15}\eta - \frac{2}{3}\right) + e_{t}^{2} \left(-\frac{64}{45}\eta + \frac{23}{15}\right)\right]}{e_{t} \left(1 - \frac{13}{6} e_{t} + \frac{7}{8} e_{t}^{2} - \frac{37}{96} e_{t}^{3}\right)} \varphi(e_{t})
\nn \\
&- \frac{8191 \pi (1 - e_{t})^{4}}{672 e_{t} \left(1 - \frac{13}{6} e_{t} + \frac{7}{8} e_{t}^{2} - \frac{37}{96} e_{t}^{3}\right)} \left[\sqrt{1 - e_{t}^{2}} \tilde{\psi}(e_{t}) - (1 - e_{t})^{2} (1 + e_{t}) \psi(e_{t})\right]
\nn \\
&- \frac{583 \pi \eta (1 - e_{t})^{4}}{24 e_{t} \left(1 - \frac{13}{6} e_{t} + \frac{7}{8} e_{t}^{2} - \frac{37}{96} e_{t}^{3}\right)}\left[\sqrt{1 - e_{t}^{2}} \tilde{\zeta}(e_{t}) - (1 - e_{t})^{2} (1 + e_{t}) \zeta(e_{t})\right] \,,
\\
V_{6}(e_{t}, \eta; v_{p}) &= \frac{116761 (1 - e_{t})^{5}}{3675 e_{t} \sqrt{1 - e_{t}^{2}} \left(1 - \frac{13}{6} e_{t} + \frac{7}{8} e_{t}^{2} - \frac{37}{96} e_{t}^{3}\right)} \left[ (1 - e_{t})^{2} (1 + e_{t}) \kappa(e_{t}) - \sqrt{1 - e_{t}^{2}} \tilde{\kappa}(e_{t})\right]
\nn \\
&+ \frac{1}{e_{t} (1 + e_{t})^{6} \sqrt{1 - e_{t}^{2}} \left(1 - \frac{13}{6} e_{t} + \frac{7}{8} e_{t}^{2} - \frac{37}{96} e_{t}^{3}\right)} \left[\frac{91271}{18900} + \sum_{k=1}^{10} {\cal{S}}_{v}^{(k)} e_{t}^{k}\right]
\nn \\
&+ \frac{1}{e_{t} (1 + e_{t})^{6} \left(1 - \frac{13}{6} e_{t} + \frac{7}{8} e_{t}^{2} - \frac{37}{96} e_{t}^{3}\right)} \left\{ \sum_{k=0}^{10} {\cal{P}}_{v}^{(k)}(\eta) e_{t}^{k} 
\right.
\nn \\
&\left.
+ {\rm ln}\left[\frac{(1-e_{t})^{3/2} (1 + e_{t})^{1/2} v_{p}}{1 + \sqrt{1 - e_{t}^{2}}}\right] {\cal{L}}_{v}(e_{t}) \right\}
\end{align}
where $[\phi(e_{t}), \psi(e_{t}), \zeta(e_{t}), \kappa(e_{t})]$ and $[\tilde{\varphi}(e_{t}), \tilde{\psi}(e_{t}), \tilde{\zeta}(e_{t}), \tilde{\kappa}(e_{t})]$ are the tail enhancement factors for the energy and angular momentum fluxes, respectively. These functions are defined in ~\cite{Arun:2007rg, Arun:2009mc}, with analytic representations as functions of $e_{t}$ given in~\cite{Loutrel:2016cdw}. The polynomials $[{\cal{P}}_{v}^{(k)}(\eta), {\cal{S}}_{v}^{(k)}(\eta), {\cal{L}}_{v}(e_{t})]$ are given as follows,
\begin{align}
{\cal{P}}_{v}^{(0)}(\eta) &= -\frac{7}{576}\eta - \frac{91271}{18900}\,,
\\
{\cal{P}}_{v}^{(1)}(\eta) &= \frac{928379951}{2851200} - \frac{1712}{35} {\rm ln}(2) + \frac{1572589}{8064}\eta^{2} - \frac{40006439}{72576}\eta + \frac{16}{3}\pi^{2} + \frac{531}{64}\pi^{2} \eta
\nn \\
&- \frac{957}{32}\eta^{3} - \frac{1712}{105}\gamma_{E}\,,
\\
{\cal{P}}_{v}^{(2)}(\eta) &= \frac{3263}{96}\eta^{3} - \frac{2984087}{8064}\eta^{2} - \frac{2218668437}{2661120} + \frac{160428925}{580608}\eta 
\nn \\
&+ \frac{72011}{630}\gamma_{E} - \frac{673}{18}\pi^{2} + \frac{72011}{210} {\rm ln}(2) - \frac{275}{128}\pi^{2} \eta\,,
\\
{\cal{P}}_{v}^{(3)}(\eta) &= -\frac{799142495}{580608}\eta + \frac{36167515}{16128}\eta^{2} - \frac{134803129}{2217600} + \frac{229}{6}\pi^{2} - \frac{24503}{70} {\rm ln}(2) 
\nn \\
&+ \frac{62183}{1536}\pi^{2} \eta - \frac{99043}{128}\eta^{3} - \frac{24503}{210}\gamma_{E}\,,
\\
{\cal{P}}_{v}^{(4)}(\eta) &= \frac{340441}{384} \eta^{3} - \frac{44827219}{8064} \eta^{2} - \frac{26356578637}{7983360} + \frac{573433099}{96768}\eta + \frac{44833}{126}\gamma_{E} 
\nn \\
&- \frac{2095}{18}\pi^{2} + \frac{44833}{42} {\rm ln}(2) + \frac{8155}{512}\pi^{2} \eta\,,
\\
{\cal{P}}_{v}^{(5)}(\eta) &= -\frac{1514002771}{193536}\eta + \frac{203612819}{21504} \eta^{2} + \frac{20453}{3072}\pi^{2} \eta + \frac{86977806011}{53222400} + \frac{109}{4}\pi^{2} 
\nn \\
&- \frac{34989}{140} {\rm ln}(2) - \frac{5262899}{1536} \eta^{3} - \frac{11663}{140}\gamma_{E}\,,
\\
{\cal{P}}_{v}^{(6)}(\eta) &= \frac{1795665}{512} \eta^{3} - \frac{113569023}{7168} \eta^{2} + \frac{3355299199}{165888}\eta - \frac{1042325420551}{106444800} + \frac{103897}{720}\gamma_{E} 
\nn \\
&- \frac{6797}{144}\pi^{2} + \frac{103897}{240} {\rm ln}(2) + \frac{260347}{12288}\pi^{2} \eta\,,
\\
{\cal{P}}_{v}^{(7)}(\eta) &= \frac{66399433}{6144} \eta^{2} - \frac{58228753}{4608}\eta - \frac{3253}{6144}\pi^{2} \eta + \frac{103183627181}{17740800} + \frac{23}{16}\pi^{2} - \frac{7383}{560} {\rm ln}(2) 
\nn \\
&- \frac{1840299}{512} \eta^{3} - \frac{2461}{560}\gamma_{E}\,,
\\
{\cal{P}}_{v}^{(8)}(\eta) &= \frac{4377905}{1536} \eta^{3} - \frac{109207321771}{23654400} - \frac{390150899}{43008} \eta^{2} + \frac{1361907947}{129024}\eta + \frac{10593}{2240}\gamma_{E} 
\nn \\
&- \frac{99}{64}\pi^{2} + \frac{31779}{2240} {\rm ln}(2) + \frac{4059}{4096}\pi^{2} \eta\,,
\\
{\cal{P}}_{v}^{(9)}(\eta) &= \frac{1559222101}{1892352} + \frac{82564327}{43008} \eta^{2} - \frac{254674141}{129024}\eta - \frac{390379}{512} \eta^{3}\,,
\\
{\cal{P}}_{v}^{(10)}(\eta) &= -\frac{2287742459}{11354112} - \frac{366721}{672}\eta^{2} + \frac{11419241}{21504}\eta + \frac{10619}{48}\eta^{3}\,,
\\
{\cal{S}}_{v}^{(1)}(\eta)&= -\frac{162703}{1440} - \frac{93}{4}\eta^{2} + \frac{42221}{336}\eta - \frac{41}{64}\pi^{2} \eta\,,
\\
{\cal{S}}_{v}^{(2)}(\eta)&= -\frac{975}{4}\eta^{2} - \frac{76684597}{151200} + \frac{138869}{112}\eta - \frac{369}{64}\pi^{2} \eta\,,
\\
{\cal{S}}_{v}^{(3)}(\eta)&= -\frac{1505201}{336}\eta + \frac{13471}{16}\eta^{2} + \frac{997437}{320} + \frac{9307}{512}\pi^{2} \eta\,,
\\
{\cal{S}}_{v}^{(4)}(\eta)&=-\frac{2727}{2}\eta^{2} - \frac{87322003}{16800} + \frac{12042805}{2016}\eta - \frac{24887}{3072}\pi^{2} \eta\,,
\\
{\cal{S}}_{v}^{(5)}(\eta)&= -\frac{1522313}{336}\eta + \frac{95255}{64}\eta^{2} - \frac{25625}{2048}\pi^{2} \eta + \frac{16734821}{3840}\,,
\\
{\cal{S}}_{v}^{(6)}(\eta)&= \frac{9333}{64}\eta^{2} - \frac{7698991}{4032}\eta + \frac{1808126359}{1209600} + \frac{39565}{3072}\pi^{2} \eta\,,
\\
{\cal{S}}_{v}^{(7)}(\eta)&= -\frac{118445}{64}\eta^{2} + \frac{39219179}{5376}\eta - \frac{10291}{2048}\pi^{2} \eta - \frac{141403667}{23040}\,,
\\
{\cal{S}}_{v}^{(8)}(\eta)&= \frac{4839771539}{1209600} + \frac{88351}{64}\eta^{2} - \frac{10136729}{2016}\eta + \frac{1517}{1536}\pi^{2} \eta\,,
\\
{\cal{S}}_{v}^{(9)}(\eta)&= -\frac{627055}{512} - \frac{14603}{32}\eta^{2} + \frac{406929}{256}\eta\,,
\\
{\cal{S}}_{v}^{(10)}(\eta)&= \frac{182905}{896} + \frac{1295}{16}\eta^{2} - \frac{123605}{448}\eta\,,
\\
{\cal{L}}_{v}(e_{t}) &= \frac{10593}{2240} e_{t}^{8} - \frac{2461}{560} e_{t}^{7} + \frac{103897}{720} e_{t}^{6} - \frac{11663}{140} e_{t}^{5} + \frac{44833}{126} e_{t}^{4} - \frac{24503}{210} e_{t}^{3} 
\nn \\
&+ \frac{72011}{630} e_{t}^{2} - \frac{1712}{105} e_{t}\,.
\end{align}
This completes the amplitude vector fields to 3PN order.

%
\bibliography{master}
\end{document}